\documentclass[reprint,superscriptaddress,amsmath,amssymb,aps,prl,longbibliography,floatfix]{revtex4-2}

\usepackage{amsmath}
\usepackage{amsfonts}
\usepackage{amssymb}
\usepackage{array}
\usepackage{bbm}
\usepackage{bm}
\usepackage{bbold}
\usepackage{braket}
\usepackage{bigints}
\usepackage{dsfont}
\usepackage[dvipsnames]{xcolor}
\usepackage{epsfig}
\usepackage{pdfpages} 
\usepackage{tikz} 

\usepackage{epstopdf}
\usepackage{pdfpages}
\usepackage{pgffor}
\usepackage{graphicx}
\usepackage{multirow}
\usepackage{mathtools}
\usepackage[T1]{fontenc}
\usepackage{newtxtext}
\usepackage[libertine]{newtxmath}
\usepackage{tensor}
\usepackage[utf8]{inputenc}
\usepackage[unicode=true,breaklinks=true,colorlinks=true]{hyperref}
\usepackage[all]{hypcap}

\hypersetup{citecolor=blue,urlcolor=blue}

\makeatletter 
\AtBeginDocument{\let\LS@rot\@undefined} 
\makeatother

\usepackage[normalem]{ulem}

\definecolor{cl1}{RGB}{255, 155, 0}
\definecolor{cl2}{RGB}{145, 82, 186}

\begin{document}

\title{Identifying possible mechanism for quantum needle in chemical magnetoreception}%

\author{Xiaoyu Chen}
\affiliation{School of Physics, Hubei Key Laboratory of Gravitation and Quantum Physics, Huazhong University of Science and Technology, Wuhan 430074, China}
\affiliation{International Joint Laboratory on Quantum Sensing and Quantum Metrology, Institute for Quantum Science and Engineering, Huazhong University of Science and Technology, Wuhan 430074, China}
\author{Haibin Liu}
\email{liuhb@hubu.edu.cn}
\affiliation{School of Physics, Hubei University, Wuhan 430074, China}
\author{Jianming Cai}
\email{jianmingcai@hust.edu.cn}
\affiliation{School of Physics, Hubei Key Laboratory of Gravitation and Quantum Physics, Huazhong University of Science and Technology, Wuhan 430074, China}
\affiliation{International Joint Laboratory on Quantum Sensing and Quantum Metrology, Institute for Quantum Science and Engineering, Huazhong University of Science and Technology, Wuhan 430074, China}

\date{\today}
             
\begin{abstract}
The radical pair mechanism is an important model that may provide a basis for biological magnetoreception. To account for the high orientation precision of the real avian compass, P. J. Hore et al. proposed an intriguing phenomenon called {\it quantum needle} [\href{https://www.pnas.org/doi/abs/10.1073/pnas.1600341113}{Proc. Natl. Acad. Sci. 113, 4634 (2016)}], where a spike-like feature emerges in the fractional yield signal. However, it is believed that quantum needle requires the radical pair lifetime to be longer than a few microseconds and thus poses stern challenges in realistic biological systems. Here, we exploit the optimization techniques and find a novel class of model system, which sustains much more prominent features of quantum needle and significantly relaxes the requirement for radical pair lifetime. Even more surprisingly, we find that the characteristics of quantum needle retain a narrow functional window around the geomagnetic field, which is absent in the previous model systems. Therefore, our work provides essential evidence for identifying the possible physical mechanism for quantum needle in chemical magnetoreception.
\end{abstract}

\maketitle

{\it Introduction.---} In recent years, quantum biology has attracted widespread attention~\cite{ball2011physics,cai2016quantum,hochstoeger2020biophysical}, which aims to explore the role of quantum effects in biological systems~\cite{cai2010quantum,hogben2012entanglement,mohseni2014quantum,huelga2013vibrations, lambert2013quantum, cao2020quantum}. A special and important example of quantum biology is the mechanism of magnetic navigation~\cite{xu2021magnetic,qin2016magnetic,maeda2008chemical}. It is well known that many species have been shown to possess geomagnetic navigation capabilities, using the geomagnetic field for orientation and migration~\cite{wiltschko2006magnetoreception,bradlaugh2023essential,bassetto2023no}. As compared with the model of magnetite-based magnetic perception~\cite{Jam_78_Sci, Jos_81_Bio, Fle_07_Natu, Hsu_07_plos, Shaw_15_magnetic}, the radical pair mechanism(RPM) ~\cite{ritz2000model, maeda2008chemical, rodgers2009chemical} involves non-trivial quantum effect and has attracted intensive interest both theoretically and experimentally~\cite{cai2013chemical,phillips1992behavioural, ritz2004resonance,mims2021readout,smith2024optimality,liu2017scheme,finkler2021quantum,aiq2022,xiao2020magnetic,PhysRevA.95.032129}. 

However, it remains mysterious why migratory birds can detect the geomagnetic field vector with such high ($<5^{\circ}$) precision~\cite{lefeldt2015migratory,aakesson2001avian}. To reveal the underlying mechanism responsible for the high orientation precision of the avian compass, it was found strikingly that an ultra-sharp dependence of the fractional yield signal on the angle of the magnetic field (called quantum needle) would appear if the radical pair lifetime exceed longer than a few microseconds, accompanied by avoided crossings of spin energy-levels~\cite{hiscock2016quantum}. Therefore, the feasibility of the phenomenon of the quantum needle in real biological systems confronts severe challenges. 
In this work, we address the critical problem, namely to what extent it is possible to relax the requirement of radical pair lifetime to observe quantum needle in chemical magnetoreception using quantum optimization. Following the theoretical framework of quantum needle~\cite{hiscock2016quantum}, we start by considering a radical pair model system involving two electron spins that each of them interacts with a nearby nuclear spin respectively, and optimize the relevant hyperfine parameters to maximize the directional precision empowered by quantum needle. Although the model system is simple, it is remarkable that quantum optimization enables us to find model systems that can improve the directional precision by an order of magnitude. 
In particular, we identify a new type of quantum needle model system, the dynamic feature of which is notably different from the conventional one~\cite{hiscock2016quantum}. Using the concept of global coherence \cite{cai2013chemical}, we reveal that the directional precision in the new model system almost solely depends on the coherent terms. In contrast, in the conventional model system~\cite{hiscock2016quantum}, quantum needle is mainly contributed by the incoherent terms, while the coherent terms partially weaken the feature of quantum needle. The result thus provides very interesting insights into the role of coherence in chemical magnetoreception. Amazingly, we find that the present new model system shows the most prominent sensitivity near $50~\mu \rm{T}$. Such behavior of a functional window around the geomagnetic field~\cite{wiltschko1972science, zvereva2010} nevertheless does not exist for the conventional system. Furthermore, the model system demonstrates significantly better noise immunity, and the phenomenon of quantum needle may still be observed under relatively short coherent time and a certain amount of noise. We expect that the present work will not only help us better understand the high directional precision of chemical magnetoreception from the perspective of quantum mechanics but also inspire biomimetic designs of novel magnetoreception devices~\cite{PhysRevLett.121.096001}.

\begin{figure*}[t]
\includegraphics[width=18cm]{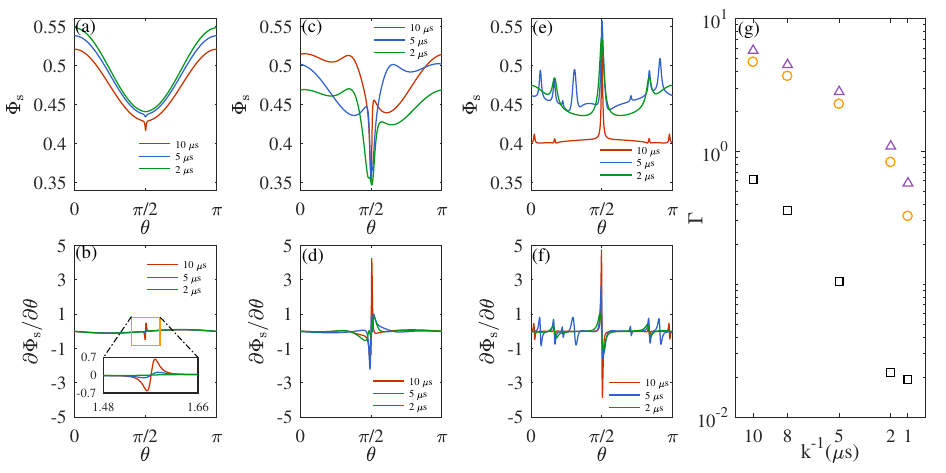}
\caption{\label{fig1}From left to right, the first three columns show that singlet yield $\Phi_s(\theta)$ and its partial derivative $\frac{\partial\Phi_s}{\partial\theta}$ as functions of the angle $\theta$ with radical pair lifetime $k^{-1}=10$, $5$, $2$~${\mu \rm{s}}$, respectively. The parameters are set to be $C=C_0=(-0.0989, -0.0989, 1.7569, 0, 0, 1.0812, 0, 0, 0)~\rm{mT}$ \cite{hiscock2016quantum} in (a-b), and the parameters (see Table \ref{table1}) for the model system I (c-d) and the model system II (e-f). The singlet yield takes the minimum value for the model system I (c), which is the same as the previous work that is demonstrated in (a), while it takes the maximum value for the model system II (e), when $\theta$ is around $90^{\circ}$. (g) compares the directional precision $\Gamma$ [see Eq.~\eqref{eq:goal}] for $C=C_0$ ($\square$), the model system I (\raisebox{-0.8pt}{\large\color{cl1}{$\circ$}}) and the model system II ({\color{cl2}{$\triangle$}}).}
\label{fig:figure1}
\end{figure*}
{\it Theoretical model and optimization methodology.---} We start by considering a simple model, in which each radical electron interacts with a nearby nucleus. The corresponding Hamiltonian is written as follows~\cite{steiner1989magnetic},
\begin{equation} \label{H}
H={I_1} \cdot \textbf{A}_1 \cdot {S_1} +{I_2} \cdot \textbf{A}_2 \cdot {S_2} +\gamma \vec{B} \cdot ({S_1}+{S_2}),
\end{equation}
where ${I_i}$ are the nuclear spin operators $(i=1,2)$, $\textbf{A}_1$ and $\textbf{A}_2$ are the hyperfine tensors, ${S_i}$ are the radical electronic spin Pauli operators $(i=1,2)$, $\gamma=\frac{1}{2}\mu_0 g$ is the electron gyromagnetic ratio with $\mu_0$ being the Bohr magneton and $g=2$ being electron Land\'{e} g-factor. We follow the assumption that the radical pair could have a large distance such that the dipole-dipole and exchange interactions can be neglected~\cite{o2005influence, efimova2008role,hiscock2016quantum}. The direction of the magnetic field is characterized by $(\theta,\phi)$, namely $\vec{B}=B_0(sin\theta cos\phi, sin\theta sin\phi, cos\theta)$, where $B_0$ is the strength of the magnetic field, $\theta$ and $\phi$ corresponds to the orientation of the magnetic field with respect to the hyperfine tensor~\cite{PhysRevLett.106.040503}. Due to symmetry, without loss of generality, we set $\phi=0$ and focus on the $\theta$ in the range $[0,\pi]$.

The radical pair is assumed to be created in a spin-correlated electronic singlet state by light, and the relevant nuclear spin is in the maximally mixed state, so the initial state is
\begin{equation}
\rho_{0}=\frac{1}{M}P^{s},
\end{equation}
where $M$ is the total number of nuclear spin configurations, $P^s=\frac{1}{4}I-{S_1} \cdot {S_2}$ is the singlet projection operator.
To simplify calculations, the singlet and triplet recombination rates in our work are taken as the same value, i.e., $k_S = k_T = k$, thus $k^{-1}$ represents the radical pair lifetime. Under this circumstance, we can use an efficient approach (see Supplementary Information~\cite{SI}) to solve the dynamics~\cite{doi:10.1080/00268979809483134}, which is governed by
\begin{equation}
\dot{\rho}=-\frac{i}{\hbar}\left[H,\rho\right]-k\rho.\label{eq:ME3}
\end{equation}
The solution of Eq.~\eqref{eq:ME3} takes the form 
\begin{equation}
\rho(t) = e^{-\frac{i}{\hbar}Ht}\rho_{0}e^{\frac{i}{\hbar}Ht}e^{-kt},
\end{equation}
and the time-dependent probability of the radical pair in a singlet state is thus given by
$p_{s}(t) = \text{Tr}[\rho(t)P^{s}]$.
Hence the singlet product yield can be written as~\cite{doi:10.1080/00268979809483134}
\begin{equation}
  \Phi_{s}  =  \int_{0}^{\infty}k p_s(t) dt=  \frac{1}{M}\sum_{m,n=1}^{4M}{\left|{P_{mn}^s}\right |}^2 \frac{k^2}{k^2+(\omega_m-\omega_n)^2},\label{eq:phis}
\end{equation}
where $P_{mn}^s$ are the matrix elements $\bra{m} P^s \ket{n}$, in which $\left|m\right\rangle $ and $\left|n\right\rangle $ are eigenstates of $H$ with energies $\hbar\omega_{m}$ and $\hbar\omega_{n}$, respectively. The 4M$\times$4M terms are grouped into coherent terms with $m\ne n$ and incoherent terms with $m=n$.

\begin{table}[t]
  \vspace{10pt}
\begin{ruledtabular}
\begin{tabular}{cccc|ccc}
 & \multicolumn{3}{c}{Model system I} & \multicolumn{3}{c}{Model system II} \\
 $k^{-1}({\mu \rm{s}})$ & $10$ & $5$ & $2$ & $10$ & $5$ & $2$
\\
\hline
$A_{xx}(\rm{mT})$ & -0.2678 & 0.3622  & 0.3710  & 0.0500  & -0.0501 & -0.0501 \\
$A_{yy}(\rm{mT})$ & 0.0001  & 0.0002  & -0.0002 & 0.0006  & 0.0000  & 0.0000   \\
$A_{zz}(\rm{mT})$ & -1.8172 & 1.7458  & -1.1338 & 0.0005  & 0.0000  & 0.0000 \\
$B_{xx}(\rm{mT})$ & 1.8366  & -1.7830 & -0.0002 & -0.0002 & -0.0300 & 0.0867  \\
$B_{yy}(\rm{mT})$ & -0.0305 & 0.0103  & 0.0001  & -0.0866 & 0.0000  & 0.0000  \\
$B_{zz}(\rm{mT})$ & -0.0039 & 0.0017  & 1.1932  & -0.0002 & -0.0058 & 0.0000 \\
$\alpha$          & -1.0042 & -0.1669 & 0.0266  & 1.2461  & -1.1853 & -0.0006 \\
$\beta$           & -1.5521 & -1.5867 & -0.0058 & -2.2840 & 1.2585  & -1.5776 \\
$\gamma$          & -2.0299 & 2.7745  & -1.3635 & -1.1465 & 0.3668  & 0.0002  \\
\end{tabular}
\end{ruledtabular}
\caption{\label{table1}
The optimized hyperfine tensors [see $\textbf{A}_1$ and $\textbf{A}_2$ in Eq.~\eqref{tensor}] for two different types of model systems with $k^{-1}=10~{\mu\rm{s}}, 5~{\mu\rm{s}}, 2~{\mu\rm{s}}$ respectively. The data with $k^{-1}=8~{\mu\rm{s}}, 1~{\mu\rm{s}}$ are demonstrated in Table SI of Supplementary Information~\cite{SI}.
}
\end{table}
Our goal is to optimize the hyperfine tensors in the above model in order to achieve an even sharper spike. There are two hyperfine tensors ($\textbf{A}_1$ and $\textbf{A}_2$) in the model, see Eq.~\eqref{H}. Here, we take the principal hyperfine axes of one nucleus as our
coordinate axes, and thus $\textbf{A}_{(1,2)}$ takes the form:
\begin{equation}
\textbf{A}_1=\left(\begin{matrix}
A_{xx} & 0 & 0 \\
0 & A_{yy} & 0 \\
0  & 0 & A_{zz} \\
\end{matrix}\right), \quad \textbf{A}_2=M\left(\begin{matrix}
B_{xx} & 0 & 0 \\
0 & B_{yy} & 0 \\
0  & 0 & B_{zz} \\
\end{matrix}\right) M^{T},
\label{tensor}
\end{equation}
where $\textbf{A}_2$ is not diagonal and has more degrees of freedom with $M=R_{z}(\gamma)R_{y}(\beta)R_{x}(\alpha)$, where $R_{x}(\alpha)$, $R_{y}(\beta)$ and $R_{z}(\gamma)$ represent rotations around the $\hat{x}(\hat{y},\hat{z})$ by angle $\alpha (\beta,\gamma)$. Overall there are nine real parameters $C=(A_{xx}, A_{yy}, A_{zz}, B_{xx}, B_{yy}, B_{zz}, \alpha, \beta, \gamma$) for optimization, and we choose the absolute value of the first order derivative of the singlet yield with respect to $\theta$(rad) to characterize the directional precision of the chemical compass. The maximum absolute value within the range of our concern ($90^{\circ}-5^{\circ}\leqslant\theta\leqslant90^{\circ}+5^{\circ}$~\cite{lefeldt2015migratory,aakesson2001avian}) is defined as the target function $\Gamma$, which takes the form
\begin{equation}\label{eq:goal}
\Gamma=\max\left(\left|\frac{\partial\Phi_{s}}{\partial\theta}\right|\right),\:\theta\in\left[\frac{\pi}{2}-\frac{5\pi}{180},\,\frac{\pi}{2}+\frac{5\pi}{180}\right].
\end{equation}

{\it New model system for quantum needle.---} Considering that the magnitude of the geomagnetic field ranges from 25 to 65$~\mu$T and the radical pair lifetime of interest is 1-10~${\mu \rm{s}}$, we set $B_0=50 ~\mu$T and $k^{-1}=10~{\mu\rm{s}}, 8~{\mu\rm{s}}, 5~{\mu\rm{s}}, 2~{\mu\rm{s}}, 1~{\mu\rm{s}}$ respectively in our optimization. Taking into account realistic factors, the range of $A_{xx}$, $A_{yy}$, $A_{zz}$, $B_{xx}$, $B_{yy}$, $B_{zz}$ is from $-2~\rm{mT}$ to $2~\rm{mT}$, and the range of $\alpha$, $\beta$, $\gamma$ is from $-\pi$ to $\pi$.
By numerical optimization, we find two different types of optimal model systems that may lead to the phenomenon of quantum needle. The corresponding optimized parameters are listed in Table \ref{table1}. Note that the optimized parameters are chosen from the best several results according to the global symmetry and the symmetry near $90^\circ$.  
For the model system I, the singlet yield shows a valley, see Fig.~\ref{fig1}(c), which is similar to the unoptimized one presented in the previous work~\cite{hiscock2016quantum} [see Fig.~\ref{fig1}(a)]. The second type of model system shows a peak [see Fig.~\ref{fig1}(e)], which represents a whole new model system with outstanding properties as presented in the following detailed analysis. With our optimized hyperfine tensors, the directional precision of quantum needle in both model systems is significantly improved. It can be seen from Fig.~\ref{fig1}(g) that the optimized precision is about ten times higher than the original one in the previous work \cite{hiscock2016quantum}, which offers the possibility to observe quantum needle with a radical pair lifetime that is an order of magnitude shorter. Thus, these results may shed light on the explanation of the behavior experiments of birds' high directional precision \cite{lefeldt2015migratory,aakesson2001avian}.
\begin{figure}[t]
\hspace{-0.4cm}\includegraphics[width=9cm]{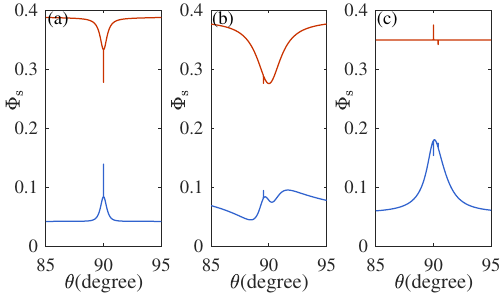}
\caption{\label{fig2} The contributions of incoherent (red) and coherent terms (blue) to quantum needles are compared for three sets of parameters: (a) $C_0=$~(-0.0989, -0.0989, 1.7569, 0, 0, 1.0812, 0, 0, 0)~$\rm{mT}$ \cite{hiscock2016quantum}, (b) $C_1=$~(-0.2678, 0.0001, -1.8172, 1.8366, -0.0305, -0.0039, -1.0042, -1.5521, -2.0299)$~\rm{mT}$ of the model system I in Table \ref{table1}, and (c) $C_2=$~(0.0500, 0.0006, 0.0005, -0.0002, -0.0866, -0.0002, 1.2461, -2.2840, -1.1465)$~\rm{mT}$ of the model system II in Table \ref{table1}. The radical pair lifetime is chosen as $k^{-1}=10$~${\mu \rm{s}}$.}
\end{figure}
To reveal the essential physics of these two different types of model systems, we group the contributions to the singlet yield according to Eq.~\eqref{eq:phis} into coherent ($m\ne n$) and incoherent terms ($m=n$) in the global basis of the total system \cite{cai2013chemical}. The results are demonstrated in Fig.~\ref{fig2}. For the model system I, the appearance of quantum needle mainly results from the incoherent terms, while the coherent terms partially weaken the feature of quantum needle, see Fig.~\ref{fig2}(a) \& (b). In contrast, for the model system II that we first find in the present work, quantum needle is solely contributed by the coherent terms without any canceling out effect, see Fig.~\ref{fig2}(c) (and see Supplementary Information \cite{SI}). It is worth pointing out that, there are very narrow paired needles that appear as vertical lines in all three examples of Fig.~\ref{fig2}, which however, are vanished when the coherent terms and the incoherent terms are added together [see Fig.~\ref{fig1}(a), (c) \& (e)]. We conjecture that this feature is also related to avoided crossing that is responsible for quantum needle~\cite{hiscock2016quantum,Bez_23_cp}. These paired narrow lines cancel each other out if the gap, i.e. the minimum difference between the energy levels involved in an avoided crossing, is far smaller than $\hbar k$ (see Supplementary Information~\cite{SI}).

\begin{figure}[t]
\hspace{-0.2cm}
\includegraphics[width=8.8cm]{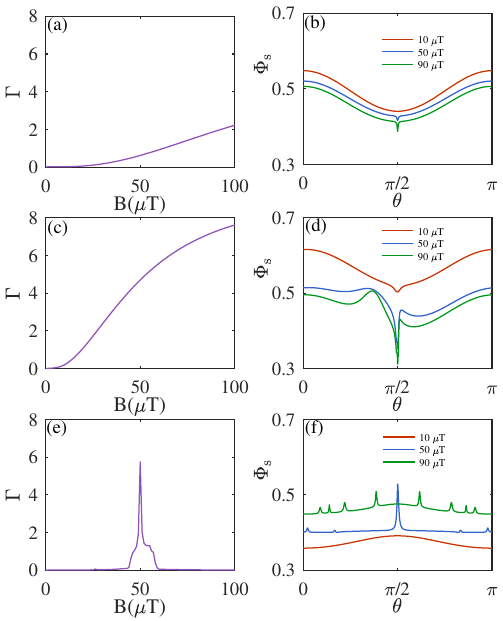}%
\caption{\label{fig3}The directional precision $\Gamma$ for the singlet yield (as defined in Eq.\ref{eq:goal}) as a function of the strength of the magnetic field $B$ in (a, c, e). (b, d, f) show the singlet yield $\Phi_s(\theta)$ for the magnetic field with $B=10,50,90~(\mu \rm{T})$, respectively. In (a, b), (c, d) and (e, f),  the parameter $C$ are set to be $C_0$, $C_1$ and $C_2$, respectively, which are the same parameters in Fig.~\ref{fig2}. The radical pair lifetime $k^{-1}$ is set as $10$~${\mu \rm{s}}$.
}
\end{figure}

{\it Functional window around the geomagnetic field.---} One essential characteristic feature of the chemical compass model of avian magnetoreception is the narrow “functional window”~\cite{wiltschko2010directional}. Behavioral experiments show that an avian compass could only work within a certain range of magnetic field strength. Once the intensity of the magnetic field increases or decreases by about 25\% of the local geomagnetic field, the avian compass would not work properly~\cite{wiltschko2006avian}. Thus, the demonstration of a functional window will serve as strong evidence for the candidate model of avian magnetoreception.
We note that for the conventional model system of quantum needle~\cite{hiscock2016quantum}, the narrow functional window phenomenon is missing, i.e. the directional precision increases as the magnetic field increases [see Fig.~\ref{fig3}(a) and (b)]. Our optimized model system I shows a similar behavior and also fails at explaining the functional window phenomenon [see Fig.~\ref{fig3}(c) and (d)]. Surprisingly, our optimized model system II exhibits high directional precision in a narrow functional window around $B=50~\mu \rm{T}$ [see Fig.~\ref{fig3}(e) and (f)], which is exactly what one would expect for an avian magnetoreception model. We remark that the width of the functional window depends on the radical pair lifetime. The shorter the radical pair lifetime that we used for optimization the wider magnetic field functional window will show up (see Supplementary Information~\cite{SI}).
{\it Influence of environmental noise.---}The avian magnetoreception occurs in a warm living system, thus environmental noise is inevitable, which will cause decoherence in quantum systems and could be detrimental to the function of quantum needle. Here, for simplicity we model the noise as pure local dephasing, thus the total system is described by the following Lindblad master equation\cite{nielsen2010quantum,PhysRevLett.106.040503,PhysRevA.85.040304},

\begin{equation}\label{eq:Lin}
\dot\rho=-\frac{i}{\hbar}\left[H,\rho\right]+\sum_{i}\left(L_i \rho L_i^{\dagger}-\frac{1}{2} (L_i^{\dagger} L_i \rho+\rho L_i^{\dagger} L_i) \right)-k\rho,
\end{equation}
where ${L}_{1}$ = ${\gamma}^{1/2} \sigma_z^{(1)}$, $L_2$ = ${\gamma}^{1/2}\sigma_z^{(2)}$, $\gamma$ is the dephasing rate. 

\begin{figure}[t]
\hspace{-0.2cm}
\includegraphics[width=8.8cm]{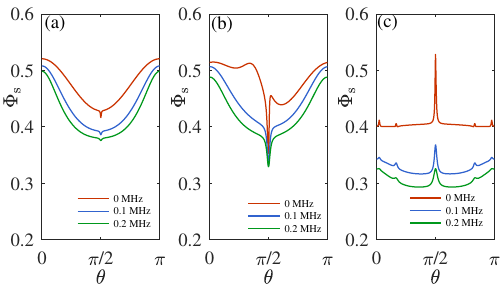}%
\caption{\label{fig4} The influence of dephasing noise on quantum needle. The red, blue and green curves correspond to the singlet yield $\Phi_s(\theta)$ for dephasing rates $\gamma=0,0.1,0.2 ~\rm{MHz}$, respectively. The parameters $C$ of (a), (b) and (c) are set to be $C_0$, $C_1$ and $C_2$ respectively, which are the same parameters in Fig.~\ref{fig2}. The radical pair lifetime $k^{-1}$ is set as $10$~${\mu \rm{s}}$.}
\end{figure}

Under this circumstance, the efficient approach Eq.~\eqref{eq:phis} is inapplicable, we numerically solve the differential equation Eq.~\eqref{eq:Lin} to analyze the impact on quantum needle. As shown in Fig.~\ref{fig4}, while the dephasing rate $\gamma$ increases, the quantum needle becomes less prominent. With the dephasing rate $\gamma=0.2 ~\rm{MHz}$, the quantum needle of the original work in Ref.~\cite{hiscock2016quantum} is almost unnoticeable, see Fig.~\ref{fig4}(a). We further analyze the performance of our optimized model systems [Fig.~\ref{fig4}(b) for the model system I and (c) for the model system II] with dephasing noise. The results show that, with the same $\gamma$, the directional precision $\Gamma$ of our optimized models is much larger, which means better noise immunity and thus would be more robust in noisy environment.
{\it Conclusion \& discussion.---}  To summarize, by optimizing the hyperfine couplings of a chemical compass with quantum needle, we manage to improve the directional precision by an order of magnitude. The significantly enhanced directional sensitivity relaxes the stringent requirement for the radical pair lifetime in order to sustain the appearance of quantum needle. We find that the new type of model system supporting the feature of quantum needle results from different dynamic characteristics and demonstrates a unique role of global coherence. More surprisingly, such a newly found model system exhibits a functional window around the geomagnetic field, which is consistent with behavior experiments of avian magnetoreception. Our results may help to resolve the contradiction between short radical pair lifetime and high-precision phenomena of the radical pair mechanism and provide important evidence for identifying the possible mechanism for chemical magnetoreception. The physics of quantum needle revealed in the present work will also bring certain inspiration for the design of the biomimetic chemical compass.
{\it Acknowledgments.---} We thank Iannis Kominis for helpful discussions. The work is supported by the National Natural Science Foundation of China (Grant No.~12161141011, 11874024), Shanghai Key Laboratory of Magnetic Resonance (East China Normal University),
%


\bibliography{cite}

\providecommand{\noopsort}[1]{}\providecommand{\singleletter}[1]{#1}%
\begin{thebibliography}{53}%
\makeatletter
\providecommand \@ifxundefined [1]{%
 \@ifx{#1\undefined}
}%
\providecommand \@ifnum [1]{%
 \ifnum #1\expandafter \@firstoftwo
 \else \expandafter \@secondoftwo
 \fi
}%
\providecommand \@ifx [1]{%
 \ifx #1\expandafter \@firstoftwo
 \else \expandafter \@secondoftwo
 \fi
}%
\providecommand \natexlab [1]{#1}%
\providecommand \enquote  [1]{``#1''}%
\providecommand \bibnamefont  [1]{#1}%
\providecommand \bibfnamefont [1]{#1}%
\providecommand \citenamefont [1]{#1}%
\providecommand \href@noop [0]{\@secondoftwo}%
\providecommand \href [0]{\begingroup \@sanitize@url \@href}%
\providecommand \@href[1]{\@@startlink{#1}\@@href}%
\providecommand \@@href[1]{\endgroup#1\@@endlink}%
\providecommand \@sanitize@url [0]{\catcode `\\12\catcode `\$12\catcode
  `\&12\catcode `\#12\catcode `\^12\catcode `\_12\catcode `\%12\relax}%
\providecommand \@@startlink[1]{}%
\providecommand \@@endlink[0]{}%
\providecommand \url  [0]{\begingroup\@sanitize@url \@url }%
\providecommand \@url [1]{\endgroup\@href {#1}{\urlprefix }}%
\providecommand \urlprefix  [0]{URL }%
\providecommand \Eprint [0]{\href }%
\providecommand \doibase [0]{https://doi.org/}%
\providecommand \selectlanguage [0]{\@gobble}%
\providecommand \bibinfo  [0]{\@secondoftwo}%
\providecommand \bibfield  [0]{\@secondoftwo}%
\providecommand \translation [1]{[#1]}%
\providecommand \BibitemOpen [0]{}%
\providecommand \bibitemStop [0]{}%
\providecommand \bibitemNoStop [0]{.\EOS\space}%
\providecommand \EOS [0]{\spacefactor3000\relax}%
\providecommand \BibitemShut  [1]{\csname bibitem#1\endcsname}%
\let\auto@bib@innerbib\@empty
\bibitem [{\citenamefont {Ball}(2011)}]{ball2011physics}%
  \BibitemOpen
  \bibfield  {author} {\bibinfo {author} {\bibfnamefont {P.}~\bibnamefont
  {Ball}},\ }\bibfield  {title} {\bibinfo {title} {{Physics of life: The dawn
  of quantum biology}},\ }\href
  {https://www.nature.com/articles/474272a#citeas} {\bibfield  {journal}
  {\bibinfo  {journal} {Nature}\ }\textbf {\bibinfo {volume} {474}},\ \bibinfo
  {pages} {272} (\bibinfo {year} {2011})}\BibitemShut {NoStop}%
\bibitem [{\citenamefont {Cai}(2016)}]{cai2016quantum}%
  \BibitemOpen
  \bibfield  {author} {\bibinfo {author} {\bibfnamefont {J.}~\bibnamefont
  {Cai}},\ }\bibfield  {title} {\bibinfo {title} {{Quantum biology: explore
  quantum dynamics in biological systems}},\ }\href
  {https://link.springer.com/article/10.1007/s11432-016-5592-y} {\bibfield
  {journal} {\bibinfo  {journal} {Science China Information Sciences}\ }\textbf
  {\bibinfo {volume} {59}},\ \bibinfo {pages} {1} (\bibinfo {year}
  {2016})}\BibitemShut {NoStop}%
\bibitem [{\citenamefont {Hochstoeger}\ \emph {et~al.}(2020)\citenamefont
  {Hochstoeger}, \citenamefont {Al~Said}, \citenamefont {Maestre},
  \citenamefont {Walter}, \citenamefont {Vilceanu}, \citenamefont {Pedron},
  \citenamefont {Cushion}, \citenamefont {Snider}, \citenamefont {Nimpf},
  \citenamefont {Nordmann} \emph {et~al.}}]{hochstoeger2020biophysical}%
  \BibitemOpen
  \bibfield  {author} {\bibinfo {author} {\bibfnamefont {T.}~\bibnamefont
  {Hochstoeger}}, \bibinfo {author} {\bibfnamefont {T.}~\bibnamefont
  {Al~Said}}, \bibinfo {author} {\bibfnamefont {D.}~\bibnamefont {Maestre}},
  \bibinfo {author} {\bibfnamefont {F.}~\bibnamefont {Walter}}, \bibinfo
  {author} {\bibfnamefont {A.}~\bibnamefont {Vilceanu}}, \bibinfo {author}
  {\bibfnamefont {M.}~\bibnamefont {Pedron}}, \bibinfo {author} {\bibfnamefont
  {T.~D.}\ \bibnamefont {Cushion}}, \bibinfo {author} {\bibfnamefont
  {W.}~\bibnamefont {Snider}}, \bibinfo {author} {\bibfnamefont
  {S.}~\bibnamefont {Nimpf}}, \bibinfo {author} {\bibfnamefont {G.~C.}\
  \bibnamefont {Nordmann}}, \emph {et~al.},\ }\bibfield  {title} {\bibinfo
  {title} {{The biophysical, molecular, and anatomical landscape of pigeon
  CRY4: A candidate light-based quantal magnetosensor}},\ }\href
  {https://www.science.org/doi/10.1126/sciadv.abb9110} {\bibfield  {journal}
  {\bibinfo  {journal} {Science Advances}\ }\textbf {\bibinfo {volume} {6}},\
  \bibinfo {pages} {eabb9110} (\bibinfo {year} {2020})}\BibitemShut {NoStop}%
\bibitem [{\citenamefont {Cai}\ \emph {et~al.}(2010)\citenamefont {Cai},
  \citenamefont {Guerreschi},\ and\ \citenamefont {Briegel}}]{cai2010quantum}%
  \BibitemOpen
  \bibfield  {author} {\bibinfo {author} {\bibfnamefont {J.}~\bibnamefont
  {Cai}}, \bibinfo {author} {\bibfnamefont {G.~G.}\ \bibnamefont
  {Guerreschi}},\ and\ \bibinfo {author} {\bibfnamefont {H.~J.}\ \bibnamefont
  {Briegel}},\ }\bibfield  {title} {\bibinfo {title} {{Quantum control and
  entanglement in a chemical compass}},\ }\href
  {https://journals.aps.org/prl/abstract/10.1103/PhysRevLett.104.220502}
  {\bibfield  {journal} {\bibinfo  {journal} {Physical Review Letters}\
  }\textbf {\bibinfo {volume} {104}},\ \bibinfo {pages} {220502} (\bibinfo
  {year} {2010})}\BibitemShut {NoStop}%
\bibitem [{\citenamefont {Hogben}\ \emph {et~al.}(2012)\citenamefont {Hogben},
  \citenamefont {Biskup},\ and\ \citenamefont {Hore}}]{hogben2012entanglement}%
  \BibitemOpen
  \bibfield  {author} {\bibinfo {author} {\bibfnamefont {H.~J.}\ \bibnamefont
  {Hogben}}, \bibinfo {author} {\bibfnamefont {T.}~\bibnamefont {Biskup}},\
  and\ \bibinfo {author} {\bibfnamefont {P.}~\bibnamefont {Hore}},\ }\bibfield
  {title} {\bibinfo {title} {{Entanglement and sources of magnetic anisotropy
  in radical pair-based avian magnetoreceptors}},\ }\href
  {https://journals.aps.org/prl/abstract/10.1103/PhysRevLett.109.220501}
  {\bibfield  {journal} {\bibinfo  {journal} {Physical Review Letters}\
  }\textbf {\bibinfo {volume} {109}},\ \bibinfo {pages} {220501} (\bibinfo
  {year} {2012})}\BibitemShut {NoStop}%
\bibitem [{\citenamefont {Mohseni}\ \emph {et~al.}(2014)\citenamefont
  {Mohseni}, \citenamefont {Omar}, \citenamefont {Engel},\ and\ \citenamefont
  {Plenio}}]{mohseni2014quantum}%
  \BibitemOpen
  \bibfield  {author} {\bibinfo {author} {\bibfnamefont {M.}~\bibnamefont
  {Mohseni}}, \bibinfo {author} {\bibfnamefont {Y.}~\bibnamefont {Omar}},
  \bibinfo {author} {\bibfnamefont {G.~S.}\ \bibnamefont {Engel}},\ and\
  \bibinfo {author} {\bibfnamefont {M.~B.}\ \bibnamefont {Plenio}},\
  }\href@noop {} {\emph {\bibinfo {title} {{Quantum effects in biology}}}}\
  (\bibinfo  {publisher} {Cambridge University Press},\ \bibinfo {year}
  {2014})\BibitemShut {NoStop}%
\bibitem [{\citenamefont {Huelga}\ and\ \citenamefont
  {Plenio}(2013)}]{huelga2013vibrations}%
  \BibitemOpen
  \bibfield  {author} {\bibinfo {author} {\bibfnamefont {S.~F.}\ \bibnamefont
  {Huelga}}\ and\ \bibinfo {author} {\bibfnamefont {M.~B.}\ \bibnamefont
  {Plenio}},\ }\bibfield  {title} {\bibinfo {title} {{Vibrations, quanta and
  biology}},\ }\href
  {https://www.tandfonline.com/doi/full/10.1080/00405000.2013.829687}
  {\bibfield  {journal} {\bibinfo  {journal} {Contemporary Physics}\ }\textbf
  {\bibinfo {volume} {54}},\ \bibinfo {pages} {181} (\bibinfo {year}
  {2013})}\BibitemShut {NoStop}%
\bibitem [{\citenamefont {Lambert}\ \emph {et~al.}(2013)\citenamefont
  {Lambert}, \citenamefont {Chen}, \citenamefont {Cheng}, \citenamefont {Li},
  \citenamefont {Chen},\ and\ \citenamefont {Nori}}]{lambert2013quantum}%
  \BibitemOpen
  \bibfield  {author} {\bibinfo {author} {\bibfnamefont {N.}~\bibnamefont
  {Lambert}}, \bibinfo {author} {\bibfnamefont {Y.-N.}\ \bibnamefont {Chen}},
  \bibinfo {author} {\bibfnamefont {Y.-C.}\ \bibnamefont {Cheng}}, \bibinfo
  {author} {\bibfnamefont {C.-M.}\ \bibnamefont {Li}}, \bibinfo {author}
  {\bibfnamefont {G.-Y.}\ \bibnamefont {Chen}},\ and\ \bibinfo {author}
  {\bibfnamefont {F.}~\bibnamefont {Nori}},\ }\bibfield  {title} {\bibinfo
  {title} {{Quantum biology}},\ }\href
  {https://www.nature.com/articles/nphys2474} {\bibfield  {journal} {\bibinfo
  {journal} {Nature Physics}\ }\textbf {\bibinfo {volume} {9}},\ \bibinfo
  {pages} {10} (\bibinfo {year} {2013})}\BibitemShut {NoStop}%
\bibitem [{\citenamefont {Cao}\ \emph {et~al.}(2020)\citenamefont {Cao},
  \citenamefont {Cogdell}, \citenamefont {Coker}, \citenamefont {Duan},
  \citenamefont {Hauer}, \citenamefont {Kleinekath{\"o}fer}, \citenamefont
  {Jansen}, \citenamefont {Man{\v{c}}al}, \citenamefont {Miller}, \citenamefont
  {Ogilvie} \emph {et~al.}}]{cao2020quantum}%
  \BibitemOpen
  \bibfield  {author} {\bibinfo {author} {\bibfnamefont {J.}~\bibnamefont
  {Cao}}, \bibinfo {author} {\bibfnamefont {R.~J.}\ \bibnamefont {Cogdell}},
  \bibinfo {author} {\bibfnamefont {D.~F.}\ \bibnamefont {Coker}}, \bibinfo
  {author} {\bibfnamefont {H.-G.}\ \bibnamefont {Duan}}, \bibinfo {author}
  {\bibfnamefont {J.}~\bibnamefont {Hauer}}, \bibinfo {author} {\bibfnamefont
  {U.}~\bibnamefont {Kleinekath{\"o}fer}}, \bibinfo {author} {\bibfnamefont
  {T.~L.}\ \bibnamefont {Jansen}}, \bibinfo {author} {\bibfnamefont
  {T.}~\bibnamefont {Man{\v{c}}al}}, \bibinfo {author} {\bibfnamefont {R.~D.}\
  \bibnamefont {Miller}}, \bibinfo {author} {\bibfnamefont {J.~P.}\
  \bibnamefont {Ogilvie}}, \emph {et~al.},\ }\bibfield  {title} {\bibinfo
  {title} {{Quantum biology revisited}},\ }\href
  {https://www.science.org/doi/full/10.1126/sciadv.aaz4888} {\bibfield
  {journal} {\bibinfo  {journal} {Science Advances}\ }\textbf {\bibinfo
  {volume} {6}},\ \bibinfo {pages} {eaaz4888} (\bibinfo {year}
  {2020})}\BibitemShut {NoStop}%
\bibitem [{\citenamefont {Xu}\ \emph {et~al.}(2021)\citenamefont {Xu},
  \citenamefont {Jarocha}, \citenamefont {Zollitsch}, \citenamefont
  {Konowalczyk}, \citenamefont {Henbest}, \citenamefont {Richert},
  \citenamefont {Golesworthy}, \citenamefont {Schmidt}, \citenamefont
  {D{\'e}jean}, \citenamefont {Sowood} \emph {et~al.}}]{xu2021magnetic}%
  \BibitemOpen
  \bibfield  {author} {\bibinfo {author} {\bibfnamefont {J.}~\bibnamefont
  {Xu}}, \bibinfo {author} {\bibfnamefont {L.~E.}\ \bibnamefont {Jarocha}},
  \bibinfo {author} {\bibfnamefont {T.}~\bibnamefont {Zollitsch}}, \bibinfo
  {author} {\bibfnamefont {M.}~\bibnamefont {Konowalczyk}}, \bibinfo {author}
  {\bibfnamefont {K.~B.}\ \bibnamefont {Henbest}}, \bibinfo {author}
  {\bibfnamefont {S.}~\bibnamefont {Richert}}, \bibinfo {author} {\bibfnamefont
  {M.~J.}\ \bibnamefont {Golesworthy}}, \bibinfo {author} {\bibfnamefont
  {J.}~\bibnamefont {Schmidt}}, \bibinfo {author} {\bibfnamefont
  {V.}~\bibnamefont {D{\'e}jean}}, \bibinfo {author} {\bibfnamefont {D.~J.}\
  \bibnamefont {Sowood}}, \emph {et~al.},\ }\bibfield  {title} {\bibinfo
  {title} {{Magnetic sensitivity of cryptochrome 4 from a migratory
  songbird}},\ }\href {https://www.nature.com/articles/s41586-021-03618-9}
  {\bibfield  {journal} {\bibinfo  {journal} {Nature}\ }\textbf {\bibinfo
  {volume} {594}},\ \bibinfo {pages} {535} (\bibinfo {year}
  {2021})}\BibitemShut {NoStop}%
\bibitem [{\citenamefont {Qin}\ \emph {et~al.}(2016)\citenamefont {Qin},
  \citenamefont {Yin}, \citenamefont {Yang}, \citenamefont {Dou}, \citenamefont
  {Liu}, \citenamefont {Zhang}, \citenamefont {Yu}, \citenamefont {Huang},
  \citenamefont {Feng}, \citenamefont {Hao} \emph {et~al.}}]{qin2016magnetic}%
  \BibitemOpen
  \bibfield  {author} {\bibinfo {author} {\bibfnamefont {S.}~\bibnamefont
  {Qin}}, \bibinfo {author} {\bibfnamefont {H.}~\bibnamefont {Yin}}, \bibinfo
  {author} {\bibfnamefont {C.}~\bibnamefont {Yang}}, \bibinfo {author}
  {\bibfnamefont {Y.}~\bibnamefont {Dou}}, \bibinfo {author} {\bibfnamefont
  {Z.}~\bibnamefont {Liu}}, \bibinfo {author} {\bibfnamefont {P.}~\bibnamefont
  {Zhang}}, \bibinfo {author} {\bibfnamefont {H.}~\bibnamefont {Yu}}, \bibinfo
  {author} {\bibfnamefont {Y.}~\bibnamefont {Huang}}, \bibinfo {author}
  {\bibfnamefont {J.}~\bibnamefont {Feng}}, \bibinfo {author} {\bibfnamefont
  {J.}~\bibnamefont {Hao}}, \emph {et~al.},\ }\bibfield  {title} {\bibinfo
  {title} {{A magnetic protein biocompass}},\ }\href
  {https://www.nature.com/articles/nmat4484} {\bibfield  {journal} {\bibinfo
  {journal} {Nature Materials}\ }\textbf {\bibinfo {volume} {15}},\ \bibinfo
  {pages} {217} (\bibinfo {year} {2016})}\BibitemShut {NoStop}%
\bibitem [{\citenamefont {Maeda}\ \emph {et~al.}(2008)\citenamefont {Maeda},
  \citenamefont {Henbest}, \citenamefont {Cintolesi}, \citenamefont {Kuprov},
  \citenamefont {Rodgers}, \citenamefont {Liddell}, \citenamefont {Gust},
  \citenamefont {Timmel},\ and\ \citenamefont {Hore}}]{maeda2008chemical}%
  \BibitemOpen
  \bibfield  {author} {\bibinfo {author} {\bibfnamefont {K.}~\bibnamefont
  {Maeda}}, \bibinfo {author} {\bibfnamefont {K.~B.}\ \bibnamefont {Henbest}},
  \bibinfo {author} {\bibfnamefont {F.}~\bibnamefont {Cintolesi}}, \bibinfo
  {author} {\bibfnamefont {I.}~\bibnamefont {Kuprov}}, \bibinfo {author}
  {\bibfnamefont {C.~T.}\ \bibnamefont {Rodgers}}, \bibinfo {author}
  {\bibfnamefont {P.~A.}\ \bibnamefont {Liddell}}, \bibinfo {author}
  {\bibfnamefont {D.}~\bibnamefont {Gust}}, \bibinfo {author} {\bibfnamefont
  {C.~R.}\ \bibnamefont {Timmel}},\ and\ \bibinfo {author} {\bibfnamefont
  {P.~J.}\ \bibnamefont {Hore}},\ }\bibfield  {title} {\bibinfo {title}
  {{Chemical compass model of avian magnetoreception}},\ }\href
  {https://www.nature.com/articles/nature06834} {\bibfield  {journal} {\bibinfo
   {journal} {Nature}\ }\textbf {\bibinfo {volume} {453}},\ \bibinfo {pages}
  {387} (\bibinfo {year} {2008})}\BibitemShut {NoStop}%
\bibitem [{\citenamefont {Wiltschko}\ and\ \citenamefont
  {Wiltschko}(2006)}]{wiltschko2006magnetoreception}%
  \BibitemOpen
  \bibfield  {author} {\bibinfo {author} {\bibfnamefont {R.}~\bibnamefont
  {Wiltschko}}\ and\ \bibinfo {author} {\bibfnamefont {W.}~\bibnamefont
  {Wiltschko}},\ }\bibfield  {title} {\bibinfo {title} {{Magnetoreception}},\
  }\href {https://onlinelibrary.wiley.com/doi/abs/10.1002/bies.20363}
  {\bibfield  {journal} {\bibinfo  {journal} {BioEssays}\ }\textbf {\bibinfo
  {volume} {28}},\ \bibinfo {pages} {157} (\bibinfo {year} {2006})}\BibitemShut
  {NoStop}%
\bibitem [{\citenamefont {Bradlaugh}\ \emph {et~al.}(2023)\citenamefont
  {Bradlaugh}, \citenamefont {Fedele}, \citenamefont {Munro}, \citenamefont
  {Hansen}, \citenamefont {Hares}, \citenamefont {Patel}, \citenamefont
  {Kyriacou}, \citenamefont {Jones}, \citenamefont {Rosato},\ and\
  \citenamefont {Baines}}]{bradlaugh2023essential}%
  \BibitemOpen
  \bibfield  {author} {\bibinfo {author} {\bibfnamefont {A.~A.}\ \bibnamefont
  {Bradlaugh}}, \bibinfo {author} {\bibfnamefont {G.}~\bibnamefont {Fedele}},
  \bibinfo {author} {\bibfnamefont {A.~L.}\ \bibnamefont {Munro}}, \bibinfo
  {author} {\bibfnamefont {C.~N.}\ \bibnamefont {Hansen}}, \bibinfo {author}
  {\bibfnamefont {J.~M.}\ \bibnamefont {Hares}}, \bibinfo {author}
  {\bibfnamefont {S.}~\bibnamefont {Patel}}, \bibinfo {author} {\bibfnamefont
  {C.~P.}\ \bibnamefont {Kyriacou}}, \bibinfo {author} {\bibfnamefont {A.~R.}\
  \bibnamefont {Jones}}, \bibinfo {author} {\bibfnamefont {E.}~\bibnamefont
  {Rosato}},\ and\ \bibinfo {author} {\bibfnamefont {R.~A.}\ \bibnamefont
  {Baines}},\ }\bibfield  {title} {\bibinfo {title} {{Essential elements of
  radical pair magnetosensitivity in Drosophila}},\ }\href
  {https://www.nature.com/articles/s41586-023-05735-z} {\bibfield  {journal}
  {\bibinfo  {journal} {Nature}\ }\textbf {\bibinfo {volume} {615}},\ \bibinfo
  {pages} {111} (\bibinfo {year} {2023})}\BibitemShut {NoStop}%
\bibitem [{\citenamefont {Bassetto}\ \emph {et~al.}(2023)\citenamefont
  {Bassetto}, \citenamefont {Reichl}, \citenamefont {Kobylkov}, \citenamefont
  {Kattnig}, \citenamefont {Winklhofer}, \citenamefont {Hore},\ and\
  \citenamefont {Mouritsen}}]{bassetto2023no}%
  \BibitemOpen
  \bibfield  {author} {\bibinfo {author} {\bibfnamefont {M.}~\bibnamefont
  {Bassetto}}, \bibinfo {author} {\bibfnamefont {T.}~\bibnamefont {Reichl}},
  \bibinfo {author} {\bibfnamefont {D.}~\bibnamefont {Kobylkov}}, \bibinfo
  {author} {\bibfnamefont {D.~R.}\ \bibnamefont {Kattnig}}, \bibinfo {author}
  {\bibfnamefont {M.}~\bibnamefont {Winklhofer}}, \bibinfo {author}
  {\bibfnamefont {P.}~\bibnamefont {Hore}},\ and\ \bibinfo {author}
  {\bibfnamefont {H.}~\bibnamefont {Mouritsen}},\ }\bibfield  {title} {\bibinfo
  {title} {{No evidence for magnetic field effects on the behaviour of
  Drosophila}},\ }\href {https://www.nature.com/articles/s41586-023-06397-7}
  {\bibfield  {journal} {\bibinfo  {journal} {Nature}\ }\textbf {\bibinfo
  {volume} {620}},\ \bibinfo {pages} {595} (\bibinfo {year}
  {2023})}\BibitemShut {NoStop}%
\bibitem [{\citenamefont {Gould}\ \emph {et~al.}(1978)\citenamefont {Gould},
  \citenamefont {Kirschvink},\ and\ \citenamefont {Deffeyes}}]{Jam_78_Sci}%
  \BibitemOpen
  \bibfield  {author} {\bibinfo {author} {\bibfnamefont {J.~L.}\ \bibnamefont
  {Gould}}, \bibinfo {author} {\bibfnamefont {J.~L.}\ \bibnamefont
  {Kirschvink}},\ and\ \bibinfo {author} {\bibfnamefont {K.~S.}\ \bibnamefont
  {Deffeyes}},\ }\bibfield  {title} {\bibinfo {title} {{Bees Have Magnetic
  Remanence}},\ }\href
  {https://www.science.org/doi/abs/10.1126/science.201.4360.1026} {\bibfield
  {journal} {\bibinfo  {journal} {Science}\ }\textbf {\bibinfo {volume}
  {201}},\ \bibinfo {pages} {1026} (\bibinfo {year} {1978})}\BibitemShut
  {NoStop}%
\bibitem [{\citenamefont {Kirschvink}\ and\ \citenamefont
  {Gould}(1981)}]{Jos_81_Bio}%
  \BibitemOpen
  \bibfield  {author} {\bibinfo {author} {\bibfnamefont {J.~L.}\ \bibnamefont
  {Kirschvink}}\ and\ \bibinfo {author} {\bibfnamefont {J.~L.}\ \bibnamefont
  {Gould}},\ }\bibfield  {title} {\bibinfo {title} {{Biogenic magnetite as a
  basis for magnetic field detection in animals}},\ }\href
  {https://doi.org/https://doi.org/10.1016/0303-2647(81)90060-5} {\bibfield
  {journal} {\bibinfo  {journal} {BioSystems}\ }\textbf {\bibinfo {volume}
  {13}},\ \bibinfo {pages} {181} (\bibinfo {year} {1981})}\BibitemShut
  {NoStop}%
\bibitem [{\citenamefont {Fleissner}\ \emph {et~al.}(2007)\citenamefont
  {Fleissner}, \citenamefont {Stahl}, \citenamefont {Thalau}, \citenamefont
  {Falkenberg},\ and\ \citenamefont {Fleissner}}]{Fle_07_Natu}%
  \BibitemOpen
  \bibfield  {author} {\bibinfo {author} {\bibfnamefont {G.}~\bibnamefont
  {Fleissner}}, \bibinfo {author} {\bibfnamefont {B.}~\bibnamefont {Stahl}},
  \bibinfo {author} {\bibfnamefont {P.}~\bibnamefont {Thalau}}, \bibinfo
  {author} {\bibfnamefont {G.}~\bibnamefont {Falkenberg}},\ and\ \bibinfo
  {author} {\bibfnamefont {G.}~\bibnamefont {Fleissner}},\ }\bibfield  {title}
  {\bibinfo {title} {{A novel concept of Fe-mineral-based magnetoreception:
  histological and physicochemical data from the upper beak of homing
  pigeons}},\ }\href {https://doi.org/10.1007/s00114-007-0236-0} {\bibfield
  {journal} {\bibinfo  {journal} {Naturwissenschaften}\ }\textbf {\bibinfo
  {volume} {94}},\ \bibinfo {pages} {631} (\bibinfo {year} {2007})}\BibitemShut
  {NoStop}%
\bibitem [{\citenamefont {Hsu}\ \emph {et~al.}(2007)\citenamefont {Hsu},
  \citenamefont {Ko}, \citenamefont {Li}, \citenamefont {Fann},\ and\
  \citenamefont {Lue}}]{Hsu_07_plos}%
  \BibitemOpen
  \bibfield  {author} {\bibinfo {author} {\bibfnamefont {C.-Y.}\ \bibnamefont
  {Hsu}}, \bibinfo {author} {\bibfnamefont {F.-Y.}\ \bibnamefont {Ko}},
  \bibinfo {author} {\bibfnamefont {C.-W.}\ \bibnamefont {Li}}, \bibinfo
  {author} {\bibfnamefont {K.}~\bibnamefont {Fann}},\ and\ \bibinfo {author}
  {\bibfnamefont {J.-T.}\ \bibnamefont {Lue}},\ }\bibfield  {title} {\bibinfo
  {title} {{Magnetoreception System in Honeybees (Apis mellifera)}},\ }\href
  {https://doi.org/10.1371/journal.pone.0000395} {\bibfield  {journal}
  {\bibinfo  {journal} {PLOS ONE}\ }\textbf {\bibinfo {volume} {2}},\ \bibinfo
  {pages} {1} (\bibinfo {year} {2007})}\BibitemShut {NoStop}%
\bibitem [{\citenamefont {Shaw}\ \emph {et~al.}(2015)\citenamefont {Shaw},
  \citenamefont {Boyd}, \citenamefont {House}, \citenamefont {Woodward},
  \citenamefont {Mathes}, \citenamefont {Cowin}, \citenamefont {Saunders},\
  and\ \citenamefont {Baer}}]{Shaw_15_magnetic}%
  \BibitemOpen
  \bibfield  {author} {\bibinfo {author} {\bibfnamefont {J.}~\bibnamefont
  {Shaw}}, \bibinfo {author} {\bibfnamefont {A.}~\bibnamefont {Boyd}}, \bibinfo
  {author} {\bibfnamefont {M.}~\bibnamefont {House}}, \bibinfo {author}
  {\bibfnamefont {R.}~\bibnamefont {Woodward}}, \bibinfo {author}
  {\bibfnamefont {F.}~\bibnamefont {Mathes}}, \bibinfo {author} {\bibfnamefont
  {G.}~\bibnamefont {Cowin}}, \bibinfo {author} {\bibfnamefont
  {M.}~\bibnamefont {Saunders}},\ and\ \bibinfo {author} {\bibfnamefont
  {B.}~\bibnamefont {Baer}},\ }\bibfield  {title} {\bibinfo {title} {{Magnetic
  particle-mediated magnetoreception}},\ }\href
  {https://doi.org/10.1098/rsif.2015.0499} {\bibfield  {journal} {\bibinfo
  {journal} {Journal of The Royal Society Interface}\ }\textbf {\bibinfo
  {volume} {12}},\ \bibinfo {pages} {20150499} (\bibinfo {year}
  {2015})}\BibitemShut {NoStop}%
\bibitem [{\citenamefont {Ritz}\ \emph {et~al.}(2000)\citenamefont {Ritz},
  \citenamefont {Adem},\ and\ \citenamefont {Schulten}}]{ritz2000model}%
  \BibitemOpen
  \bibfield  {author} {\bibinfo {author} {\bibfnamefont {T.}~\bibnamefont
  {Ritz}}, \bibinfo {author} {\bibfnamefont {S.}~\bibnamefont {Adem}},\ and\
  \bibinfo {author} {\bibfnamefont {K.}~\bibnamefont {Schulten}},\ }\bibfield
  {title} {\bibinfo {title} {{A model for photoreceptor-based magnetoreception
  in birds}},\ }\href {https://www.cell.com/fulltext/S0006-3495%2800%2976629-X}
  {\bibfield  {journal} {\bibinfo  {journal} {Biophysical Journal}\ }\textbf
  {\bibinfo {volume} {78}},\ \bibinfo {pages} {707} (\bibinfo {year}
  {2000})}\BibitemShut {NoStop}%
\bibitem [{\citenamefont {Rodgers}\ and\ \citenamefont
  {Hore}(2009)}]{rodgers2009chemical}%
  \BibitemOpen
  \bibfield  {author} {\bibinfo {author} {\bibfnamefont {C.~T.}\ \bibnamefont
  {Rodgers}}\ and\ \bibinfo {author} {\bibfnamefont {P.~J.}\ \bibnamefont
  {Hore}},\ }\bibfield  {title} {\bibinfo {title} {{Chemical magnetoreception
  in birds: the radical pair mechanism}},\ }\href
  {https://www.pnas.org/doi/abs/10.1073/pnas.0711968106} {\bibfield  {journal}
  {\bibinfo  {journal} {Proceedings of the National Academy of Sciences}\
  }\textbf {\bibinfo {volume} {106}},\ \bibinfo {pages} {353} (\bibinfo {year}
  {2009})}\BibitemShut {NoStop}%
\bibitem [{\citenamefont {Cai}\ and\ \citenamefont
  {Plenio}(2013)}]{cai2013chemical}%
  \BibitemOpen
  \bibfield  {author} {\bibinfo {author} {\bibfnamefont {J.}~\bibnamefont
  {Cai}}\ and\ \bibinfo {author} {\bibfnamefont {M.~B.}\ \bibnamefont
  {Plenio}},\ }\bibfield  {title} {\bibinfo {title} {{Chemical compass model
  for avian magnetoreception as a quantum coherent device}},\ }\href
  {https://journals.aps.org/prl/abstract/10.1103/PhysRevLett.111.230503}
  {\bibfield  {journal} {\bibinfo  {journal} {Physical Review Letters}\
  }\textbf {\bibinfo {volume} {111}},\ \bibinfo {pages} {230503} (\bibinfo
  {year} {2013})}\BibitemShut {NoStop}%
\bibitem [{\citenamefont {Phillips}\ and\ \citenamefont
  {Borland}(1992)}]{phillips1992behavioural}%
  \BibitemOpen
  \bibfield  {author} {\bibinfo {author} {\bibfnamefont {J.~B.}\ \bibnamefont
  {Phillips}}\ and\ \bibinfo {author} {\bibfnamefont {S.~C.}\ \bibnamefont
  {Borland}},\ }\bibfield  {title} {\bibinfo {title} {{Behavioural evidence for
  use of a light-dependent magnetoreception mechanism by a vertebrate}},\
  }\href {https://www.nature.com/articles/359142a0} {\bibfield  {journal}
  {\bibinfo  {journal} {Nature}\ }\textbf {\bibinfo {volume} {359}},\ \bibinfo
  {pages} {142} (\bibinfo {year} {1992})}\BibitemShut {NoStop}%
\bibitem [{\citenamefont {Ritz}\ \emph {et~al.}(2004)\citenamefont {Ritz},
  \citenamefont {Thalau}, \citenamefont {Phillips}, \citenamefont {Wiltschko},\
  and\ \citenamefont {Wiltschko}}]{ritz2004resonance}%
  \BibitemOpen
  \bibfield  {author} {\bibinfo {author} {\bibfnamefont {T.}~\bibnamefont
  {Ritz}}, \bibinfo {author} {\bibfnamefont {P.}~\bibnamefont {Thalau}},
  \bibinfo {author} {\bibfnamefont {J.~B.}\ \bibnamefont {Phillips}}, \bibinfo
  {author} {\bibfnamefont {R.}~\bibnamefont {Wiltschko}},\ and\ \bibinfo
  {author} {\bibfnamefont {W.}~\bibnamefont {Wiltschko}},\ }\bibfield  {title}
  {\bibinfo {title} {{Resonance effects indicate a radical-pair mechanism for
  avian magnetic compass}},\ }\href
  {https://www.nature.com/articles/nature02534} {\bibfield  {journal} {\bibinfo
   {journal} {Nature}\ }\textbf {\bibinfo {volume} {429}},\ \bibinfo {pages}
  {177} (\bibinfo {year} {2004})}\BibitemShut {NoStop}%
\bibitem [{\citenamefont {Mims}\ \emph {et~al.}(2021)\citenamefont {Mims},
  \citenamefont {Herpich}, \citenamefont {Lukzen}, \citenamefont {Steiner},\
  and\ \citenamefont {Lambert}}]{mims2021readout}%
  \BibitemOpen
  \bibfield  {author} {\bibinfo {author} {\bibfnamefont {D.}~\bibnamefont
  {Mims}}, \bibinfo {author} {\bibfnamefont {J.}~\bibnamefont {Herpich}},
  \bibinfo {author} {\bibfnamefont {N.~N.}\ \bibnamefont {Lukzen}}, \bibinfo
  {author} {\bibfnamefont {U.~E.}\ \bibnamefont {Steiner}},\ and\ \bibinfo
  {author} {\bibfnamefont {C.}~\bibnamefont {Lambert}},\ }\bibfield  {title}
  {\bibinfo {title} {{Readout of spin quantum beats in a charge-separated
  radical pair by pump-push spectroscopy}},\ }\href
  {https://www.science.org/doi/10.1126/science.abl4254} {\bibfield  {journal}
  {\bibinfo  {journal} {Science}\ }\textbf {\bibinfo {volume} {374}},\ \bibinfo
  {pages} {1470} (\bibinfo {year} {2021})}\BibitemShut {NoStop}%
\bibitem [{\citenamefont {Smith}\ \emph {et~al.}(2024)\citenamefont {Smith},
  \citenamefont {Glatthard}, \citenamefont {Chowdhury},\ and\ \citenamefont
  {Kattnig}}]{smith2024optimality}%
  \BibitemOpen
  \bibfield  {author} {\bibinfo {author} {\bibfnamefont {L.~D.}\ \bibnamefont
  {Smith}}, \bibinfo {author} {\bibfnamefont {J.}~\bibnamefont {Glatthard}},
  \bibinfo {author} {\bibfnamefont {F.~T.}\ \bibnamefont {Chowdhury}},\ and\
  \bibinfo {author} {\bibfnamefont {D.~R.}\ \bibnamefont {Kattnig}},\
  }\bibfield  {title} {\bibinfo {title} {{On the optimality of the radical-pair
  quantum compass}},\ }\href {https://arxiv.org/abs/2401.02923} {\bibfield
  {journal} {\bibinfo  {journal} {arXiv:2401.02923}\ } (\bibinfo {year}
  {2024})}\BibitemShut {NoStop}%
\bibitem [{\citenamefont {Liu}\ \emph {et~al.}(2017)\citenamefont {Liu},
  \citenamefont {Plenio},\ and\ \citenamefont {Cai}}]{liu2017scheme}%
  \BibitemOpen
  \bibfield  {author} {\bibinfo {author} {\bibfnamefont {H.}~\bibnamefont
  {Liu}}, \bibinfo {author} {\bibfnamefont {M.~B.}\ \bibnamefont {Plenio}},\
  and\ \bibinfo {author} {\bibfnamefont {J.}~\bibnamefont {Cai}},\ }\bibfield
  {title} {\bibinfo {title} {{Scheme for detection of single-molecule radical
  pair reaction using spin in diamond}},\ }\href
  {https://journals.aps.org/prl/abstract/10.1103/PhysRevLett.118.200402}
  {\bibfield  {journal} {\bibinfo  {journal} {Physical Review Letters}\
  }\textbf {\bibinfo {volume} {118}},\ \bibinfo {pages} {200402} (\bibinfo
  {year} {2017})}\BibitemShut {NoStop}%
\bibitem [{\citenamefont {Finkler}\ and\ \citenamefont
  {Dasari}(2021)}]{finkler2021quantum}%
  \BibitemOpen
  \bibfield  {author} {\bibinfo {author} {\bibfnamefont {A.}~\bibnamefont
  {Finkler}}\ and\ \bibinfo {author} {\bibfnamefont {D.}~\bibnamefont
  {Dasari}},\ }\bibfield  {title} {\bibinfo {title} {{Quantum sensing and
  control of spin-state dynamics in the radical-pair mechanism}},\ }\href
  {https://journals.aps.org/prapplied/abstract/10.1103/PhysRevApplied.15.034066}
  {\bibfield  {journal} {\bibinfo  {journal} {Physical Review Applied}\
  }\textbf {\bibinfo {volume} {15}},\ \bibinfo {pages} {034066} (\bibinfo
  {year} {2021})}\BibitemShut {NoStop}%
\bibitem [{\citenamefont {Wu}\ \emph {et~al.}(2022)\citenamefont {Wu},
  \citenamefont {Hu}, \citenamefont {Zhu}, \citenamefont {Deng},\ and\
  \citenamefont {Ai}}]{aiq2022}%
  \BibitemOpen
  \bibfield  {author} {\bibinfo {author} {\bibfnamefont {J.-Y.}\ \bibnamefont
  {Wu}}, \bibinfo {author} {\bibfnamefont {X.-Y.}\ \bibnamefont {Hu}}, \bibinfo
  {author} {\bibfnamefont {H.-Y.}\ \bibnamefont {Zhu}}, \bibinfo {author}
  {\bibfnamefont {R.-Q.}\ \bibnamefont {Deng}},\ and\ \bibinfo {author}
  {\bibfnamefont {Q.}~\bibnamefont {Ai}},\ }\bibfield  {title} {\bibinfo
  {title} {{A Bionic Compass Based on Multiradicals}},\ }\href
  {https://doi.org/10.1021/acs.jpcb.2c02711} {\bibfield  {journal} {\bibinfo
  {journal} {The Journal of Physical Chemistry B}\ }\textbf {\bibinfo {volume}
  {126}},\ \bibinfo {pages} {10327} (\bibinfo {year} {2022})}\BibitemShut
  {NoStop}%
\bibitem [{\citenamefont {Xiao}\ \emph {et~al.}(2020)\citenamefont {Xiao},
  \citenamefont {Hu}, \citenamefont {Cai},\ and\ \citenamefont
  {Zhao}}]{xiao2020magnetic}%
  \BibitemOpen
  \bibfield  {author} {\bibinfo {author} {\bibfnamefont {D.-W.}\ \bibnamefont
  {Xiao}}, \bibinfo {author} {\bibfnamefont {W.-H.}\ \bibnamefont {Hu}},
  \bibinfo {author} {\bibfnamefont {Y.}~\bibnamefont {Cai}},\ and\ \bibinfo
  {author} {\bibfnamefont {N.}~\bibnamefont {Zhao}},\ }\bibfield  {title}
  {\bibinfo {title} {{Magnetic noise enabled biocompass}},\ }\href
  {https://journals.aps.org/prl/abstract/10.1103/PhysRevLett.124.128101}
  {\bibfield  {journal} {\bibinfo  {journal} {Physical Review Letters}\
  }\textbf {\bibinfo {volume} {124}},\ \bibinfo {pages} {128101} (\bibinfo
  {year} {2020})}\BibitemShut {NoStop}%
\bibitem [{\citenamefont {Vitalis}\ and\ \citenamefont
  {Kominis}(2017)}]{PhysRevA.95.032129}%
  \BibitemOpen
  \bibfield  {author} {\bibinfo {author} {\bibfnamefont {K.~M.}\ \bibnamefont
  {Vitalis}}\ and\ \bibinfo {author} {\bibfnamefont {I.~K.}\ \bibnamefont
  {Kominis}},\ }\bibfield  {title} {\bibinfo {title} {{Quantum-limited
  biochemical magnetometers designed using the Fisher information and quantum
  reaction control}},\ }\href {https://doi.org/10.1103/PhysRevA.95.032129}
  {\bibfield  {journal} {\bibinfo  {journal} {Physical Review A}\ }\textbf
  {\bibinfo {volume} {95}},\ \bibinfo {pages} {032129} (\bibinfo {year}
  {2017})}\BibitemShut {NoStop}%
\bibitem [{\citenamefont {Lefeldt}\ \emph {et~al.}(2015)\citenamefont
  {Lefeldt}, \citenamefont {Dreyer}, \citenamefont {Schneider}, \citenamefont
  {Steenken},\ and\ \citenamefont {Mouritsen}}]{lefeldt2015migratory}%
  \BibitemOpen
  \bibfield  {author} {\bibinfo {author} {\bibfnamefont {N.}~\bibnamefont
  {Lefeldt}}, \bibinfo {author} {\bibfnamefont {D.}~\bibnamefont {Dreyer}},
  \bibinfo {author} {\bibfnamefont {N.-L.}\ \bibnamefont {Schneider}}, \bibinfo
  {author} {\bibfnamefont {F.}~\bibnamefont {Steenken}},\ and\ \bibinfo
  {author} {\bibfnamefont {H.}~\bibnamefont {Mouritsen}},\ }\bibfield  {title}
  {\bibinfo {title} {{Migratory blackcaps tested in Emlen funnels can orient at
  85 degrees but not at 88 degrees magnetic inclination}},\ }\href
  {https://journals.biologists.com/jeb/article/218/2/206/14271/Migratory-blackcaps-tested-in-Emlen-funnels-can}
  {\bibfield  {journal} {\bibinfo  {journal} {Journal of Experimental Biology}\
  }\textbf {\bibinfo {volume} {218}},\ \bibinfo {pages} {206} (\bibinfo {year}
  {2015})}\BibitemShut {NoStop}%
\bibitem [{\citenamefont {{\AA}kesson}\ \emph {et~al.}(2001)\citenamefont
  {{\AA}kesson}, \citenamefont {Morin}, \citenamefont {Muheim},\ and\
  \citenamefont {Ottosson}}]{aakesson2001avian}%
  \BibitemOpen
  \bibfield  {author} {\bibinfo {author} {\bibfnamefont {S.}~\bibnamefont
  {{\AA}kesson}}, \bibinfo {author} {\bibfnamefont {J.}~\bibnamefont {Morin}},
  \bibinfo {author} {\bibfnamefont {R.}~\bibnamefont {Muheim}},\ and\ \bibinfo
  {author} {\bibfnamefont {U.}~\bibnamefont {Ottosson}},\ }\bibfield  {title}
  {\bibinfo {title} {{Avian orientation at steep angles of inclination:
  experiments with migratory white--crowned sparrows at the magnetic North
  Pole}},\ }\href {http://doi.org/10.1098/rspb.2001.1736} {\bibfield  {journal}
  {\bibinfo  {journal} {Proceedings of the Royal Society B}\ }\textbf {\bibinfo
  {volume} {268}},\ \bibinfo {pages} {1907} (\bibinfo {year}
  {2001})}\BibitemShut {NoStop}%
\bibitem [{\citenamefont {Hiscock}\ \emph {et~al.}(2016)\citenamefont
  {Hiscock}, \citenamefont {Worster}, \citenamefont {Kattnig}, \citenamefont
  {Steers}, \citenamefont {Jin}, \citenamefont {Manolopoulos}, \citenamefont
  {Mouritsen},\ and\ \citenamefont {Hore}}]{hiscock2016quantum}%
  \BibitemOpen
  \bibfield  {author} {\bibinfo {author} {\bibfnamefont {H.~G.}\ \bibnamefont
  {Hiscock}}, \bibinfo {author} {\bibfnamefont {S.}~\bibnamefont {Worster}},
  \bibinfo {author} {\bibfnamefont {D.~R.}\ \bibnamefont {Kattnig}}, \bibinfo
  {author} {\bibfnamefont {C.}~\bibnamefont {Steers}}, \bibinfo {author}
  {\bibfnamefont {Y.}~\bibnamefont {Jin}}, \bibinfo {author} {\bibfnamefont
  {D.~E.}\ \bibnamefont {Manolopoulos}}, \bibinfo {author} {\bibfnamefont
  {H.}~\bibnamefont {Mouritsen}},\ and\ \bibinfo {author} {\bibfnamefont
  {P.~J.}\ \bibnamefont {Hore}},\ }\bibfield  {title} {\bibinfo {title} {{The
  quantum needle of the avian magnetic compass}},\ }\href
  {https://www.pnas.org/doi/abs/10.1073/pnas.1600341113} {\bibfield  {journal}
  {\bibinfo  {journal} {Proceedings of the National Academy of Sciences}\
  }\textbf {\bibinfo {volume} {113}},\ \bibinfo {pages} {4634} (\bibinfo {year}
  {2016})}\BibitemShut {NoStop}%
\bibitem [{\citenamefont {Wiltschko}\ and\ \citenamefont
  {Wiltschko}(1972)}]{wiltschko1972science}%
  \BibitemOpen
  \bibfield  {author} {\bibinfo {author} {\bibfnamefont {W.}~\bibnamefont
  {Wiltschko}}\ and\ \bibinfo {author} {\bibfnamefont {R.}~\bibnamefont
  {Wiltschko}},\ }\bibfield  {title} {\bibinfo {title} {{Magnetic Compass of
  European Robins}},\ }\href
  {https://www.science.org/doi/abs/10.1126/science.176.4030.62} {\bibfield
  {journal} {\bibinfo  {journal} {Science}\ }\textbf {\bibinfo {volume}
  {176}},\ \bibinfo {pages} {62} (\bibinfo {year} {1972})}\BibitemShut
  {NoStop}%
\bibitem [{\citenamefont {Finlay}\ \emph {et~al.}(2010)\citenamefont {Finlay},
  \citenamefont {Maus}, \citenamefont {Beggan}, \citenamefont {Bondar},
  \citenamefont {Chambodut}, \citenamefont {Chernova}, \citenamefont
  {Chulliat}, \citenamefont {Golovkov}, \citenamefont {Hamilton}, \citenamefont
  {Hamoudi}, \citenamefont {Holme}, \citenamefont {Hulot}, \citenamefont
  {Kuang}, \citenamefont {Langlais}, \citenamefont {Lesur}, \citenamefont
  {Lowes}, \citenamefont {Lühr}, \citenamefont {Macmillan}, \citenamefont
  {Mandea}, \citenamefont {McLean}, \citenamefont {Manoj}, \citenamefont
  {Menvielle}, \citenamefont {Michaelis}, \citenamefont {Olsen}, \citenamefont
  {Rauberg}, \citenamefont {Rother}, \citenamefont {Sabaka}, \citenamefont
  {Tangborn}, \citenamefont {Tøffner-Clausen}, \citenamefont {Thébault},
  \citenamefont {Thomson}, \citenamefont {Wardinski}, \citenamefont {Wei},\
  and\ \citenamefont {Zvereva}}]{zvereva2010}%
  \BibitemOpen
  \bibfield  {author} {\bibinfo {author} {\bibfnamefont {C.~C.}\ \bibnamefont
  {Finlay}}, \bibinfo {author} {\bibfnamefont {S.}~\bibnamefont {Maus}},
  \bibinfo {author} {\bibfnamefont {C.~D.}\ \bibnamefont {Beggan}}, \bibinfo
  {author} {\bibfnamefont {T.~N.}\ \bibnamefont {Bondar}}, \bibinfo {author}
  {\bibfnamefont {A.}~\bibnamefont {Chambodut}}, \bibinfo {author}
  {\bibfnamefont {T.~A.}\ \bibnamefont {Chernova}}, \bibinfo {author}
  {\bibfnamefont {A.}~\bibnamefont {Chulliat}}, \bibinfo {author}
  {\bibfnamefont {V.~P.}\ \bibnamefont {Golovkov}}, \bibinfo {author}
  {\bibfnamefont {B.}~\bibnamefont {Hamilton}}, \bibinfo {author}
  {\bibfnamefont {M.}~\bibnamefont {Hamoudi}}, \bibinfo {author} {\bibfnamefont
  {R.}~\bibnamefont {Holme}}, \bibinfo {author} {\bibfnamefont
  {G.}~\bibnamefont {Hulot}}, \bibinfo {author} {\bibfnamefont
  {W.}~\bibnamefont {Kuang}}, \bibinfo {author} {\bibfnamefont
  {B.}~\bibnamefont {Langlais}}, \bibinfo {author} {\bibfnamefont
  {V.}~\bibnamefont {Lesur}}, \bibinfo {author} {\bibfnamefont {F.~J.}\
  \bibnamefont {Lowes}}, \bibinfo {author} {\bibfnamefont {H.}~\bibnamefont
  {Lühr}}, \bibinfo {author} {\bibfnamefont {S.}~\bibnamefont {Macmillan}},
  \bibinfo {author} {\bibfnamefont {M.}~\bibnamefont {Mandea}}, \bibinfo
  {author} {\bibfnamefont {S.}~\bibnamefont {McLean}}, \bibinfo {author}
  {\bibfnamefont {C.}~\bibnamefont {Manoj}}, \bibinfo {author} {\bibfnamefont
  {M.}~\bibnamefont {Menvielle}}, \bibinfo {author} {\bibfnamefont
  {I.}~\bibnamefont {Michaelis}}, \bibinfo {author} {\bibfnamefont
  {N.}~\bibnamefont {Olsen}}, \bibinfo {author} {\bibfnamefont
  {J.}~\bibnamefont {Rauberg}}, \bibinfo {author} {\bibfnamefont
  {M.}~\bibnamefont {Rother}}, \bibinfo {author} {\bibfnamefont {T.~J.}\
  \bibnamefont {Sabaka}}, \bibinfo {author} {\bibfnamefont {A.}~\bibnamefont
  {Tangborn}}, \bibinfo {author} {\bibfnamefont {L.}~\bibnamefont
  {Tøffner-Clausen}}, \bibinfo {author} {\bibfnamefont {E.}~\bibnamefont
  {Thébault}}, \bibinfo {author} {\bibfnamefont {A.~W.~P.}\ \bibnamefont
  {Thomson}}, \bibinfo {author} {\bibfnamefont {I.}~\bibnamefont {Wardinski}},
  \bibinfo {author} {\bibfnamefont {Z.}~\bibnamefont {Wei}},\ and\ \bibinfo
  {author} {\bibfnamefont {T.~I.}\ \bibnamefont {Zvereva}},\ }\bibfield
  {title} {\bibinfo {title} {{International Geomagnetic Reference Field: the
  eleventh generation}},\ }\href
  {https://doi.org/10.1111/j.1365-246X.2010.04804.x} {\bibfield  {journal}
  {\bibinfo  {journal} {Geophysical Journal International}\ }\textbf {\bibinfo
  {volume} {183}},\ \bibinfo {pages} {1216} (\bibinfo {year}
  {2010})}\BibitemShut {NoStop}%
\bibitem [{\citenamefont {Keens}\ \emph {et~al.}(2018)\citenamefont {Keens},
  \citenamefont {Bedkihal},\ and\ \citenamefont
  {Kattnig}}]{PhysRevLett.121.096001}%
  \BibitemOpen
  \bibfield  {author} {\bibinfo {author} {\bibfnamefont {R.~H.}\ \bibnamefont
  {Keens}}, \bibinfo {author} {\bibfnamefont {S.}~\bibnamefont {Bedkihal}},\
  and\ \bibinfo {author} {\bibfnamefont {D.~R.}\ \bibnamefont {Kattnig}},\
  }\bibfield  {title} {\bibinfo {title} {{Magnetosensitivity in Dipolarly
  Coupled Three-Spin Systems}},\ }\href
  {https://doi.org/10.1103/PhysRevLett.121.096001} {\bibfield  {journal}
  {\bibinfo  {journal} {Physical Review Letters}\ }\textbf {\bibinfo {volume}
  {121}},\ \bibinfo {pages} {096001} (\bibinfo {year} {2018})}\BibitemShut
  {NoStop}%
\bibitem [{\citenamefont {Steiner}\ and\ \citenamefont
  {Ulrich}(1989)}]{steiner1989magnetic}%
  \BibitemOpen
  \bibfield  {author} {\bibinfo {author} {\bibfnamefont {U.~E.}\ \bibnamefont
  {Steiner}}\ and\ \bibinfo {author} {\bibfnamefont {T.}~\bibnamefont
  {Ulrich}},\ }\bibfield  {title} {\bibinfo {title} {{Magnetic field effects in
  chemical kinetics and related phenomena}},\ }\href
  {https://doi.org/10.1021/cr00091a003} {\bibfield  {journal} {\bibinfo
  {journal} {Chemical Reviews}\ }\textbf {\bibinfo {volume} {89}},\ \bibinfo
  {pages} {51} (\bibinfo {year} {1989})}\BibitemShut {NoStop}%
\bibitem [{\citenamefont {O'Dea}\ \emph {et~al.}(2005)\citenamefont {O'Dea},
  \citenamefont {Curtis}, \citenamefont {Green}, \citenamefont {Timmel},\ and\
  \citenamefont {Hore}}]{o2005influence}%
  \BibitemOpen
  \bibfield  {author} {\bibinfo {author} {\bibfnamefont {A.~R.}\ \bibnamefont
  {O'Dea}}, \bibinfo {author} {\bibfnamefont {A.~F.}\ \bibnamefont {Curtis}},
  \bibinfo {author} {\bibfnamefont {N.~J.}\ \bibnamefont {Green}}, \bibinfo
  {author} {\bibfnamefont {C.~R.}\ \bibnamefont {Timmel}},\ and\ \bibinfo
  {author} {\bibfnamefont {P.}~\bibnamefont {Hore}},\ }\bibfield  {title}
  {\bibinfo {title} {{Influence of dipolar interactions on radical pair
  recombination reactions subject to weak magnetic fields}},\ }\href
  {https://pubs.acs.org/doi/abs/10.1021/jp0456943} {\bibfield  {journal}
  {\bibinfo  {journal} {The Journal of Physical Chemistry A}\ }\textbf
  {\bibinfo {volume} {109}},\ \bibinfo {pages} {869} (\bibinfo {year}
  {2005})}\BibitemShut {NoStop}%
\bibitem [{\citenamefont {Efimova}\ and\ \citenamefont
  {Hore}(2008)}]{efimova2008role}%
  \BibitemOpen
  \bibfield  {author} {\bibinfo {author} {\bibfnamefont {O.}~\bibnamefont
  {Efimova}}\ and\ \bibinfo {author} {\bibfnamefont {P.}~\bibnamefont {Hore}},\
  }\bibfield  {title} {\bibinfo {title} {{Role of exchange and dipolar
  interactions in the radical pair model of the avian magnetic compass}},\
  }\href {https://www.cell.com/fulltext/S0006-3495(08)70595-2} {\bibfield
  {journal} {\bibinfo  {journal} {Biophysical Journal}\ }\textbf {\bibinfo
  {volume} {94}},\ \bibinfo {pages} {1565} (\bibinfo {year}
  {2008})}\BibitemShut {NoStop}%
\bibitem [{\citenamefont {Gauger}\ \emph {et~al.}(2011)\citenamefont {Gauger},
  \citenamefont {Rieper}, \citenamefont {Morton}, \citenamefont {Benjamin},\
  and\ \citenamefont {Vedral}}]{PhysRevLett.106.040503}%
  \BibitemOpen
  \bibfield  {author} {\bibinfo {author} {\bibfnamefont {E.~M.}\ \bibnamefont
  {Gauger}}, \bibinfo {author} {\bibfnamefont {E.}~\bibnamefont {Rieper}},
  \bibinfo {author} {\bibfnamefont {J.~J.~L.}\ \bibnamefont {Morton}}, \bibinfo
  {author} {\bibfnamefont {S.~C.}\ \bibnamefont {Benjamin}},\ and\ \bibinfo
  {author} {\bibfnamefont {V.}~\bibnamefont {Vedral}},\ }\bibfield  {title}
  {\bibinfo {title} {{Sustained Quantum Coherence and Entanglement in the Avian
  Compass}},\ }\href {https://doi.org/10.1103/PhysRevLett.106.040503}
  {\bibfield  {journal} {\bibinfo  {journal} {Physical Review Letters}\
  }\textbf {\bibinfo {volume} {106}},\ \bibinfo {pages} {040503} (\bibinfo
  {year} {2011})}\BibitemShut {NoStop}%
\bibitem [{SI()}]{SI}%
  \BibitemOpen
  \href@noop {} {\bibinfo {title} {{See Supplemental Material} at
  \href{https://TBD}{TBD} for details, which includes
  \cite{schulten1978biomagnetic,ritz2000model,johnsen2005physics,rodgers2009chemical,kominis2009quantum,jones2010spin,doi:10.1080/00268979809483134,hiscock2016quantum}}}\BibitemShut
  {NoStop}%
\bibitem [{\citenamefont {Timmel}\ \emph {et~al.}(1998)\citenamefont {Timmel},
  \citenamefont {Till}, \citenamefont {Brocklehurst}, \citenamefont
  {Mclauchlan},\ and\ \citenamefont {Hore}}]{doi:10.1080/00268979809483134}%
  \BibitemOpen
  \bibfield  {author} {\bibinfo {author} {\bibfnamefont {C.}~\bibnamefont
  {Timmel}}, \bibinfo {author} {\bibfnamefont {U.}~\bibnamefont {Till}},
  \bibinfo {author} {\bibfnamefont {B.}~\bibnamefont {Brocklehurst}}, \bibinfo
  {author} {\bibfnamefont {K.}~\bibnamefont {Mclauchlan}},\ and\ \bibinfo
  {author} {\bibfnamefont {P.}~\bibnamefont {Hore}},\ }\bibfield  {title}
  {\bibinfo {title} {{Effects of weak magnetic fields on free radical
  recombination reactions}},\ }\href
  {https://doi.org/10.1080/00268979809483134} {\bibfield  {journal} {\bibinfo
  {journal} {Molecular Physics}\ }\textbf {\bibinfo {volume} {95}},\ \bibinfo
  {pages} {71} (\bibinfo {year} {1998})}\BibitemShut {NoStop}%
\bibitem [{\citenamefont {Bezchastnov}\ and\ \citenamefont
  {Domratcheva}(2023)}]{Bez_23_cp}%
  \BibitemOpen
  \bibfield  {author} {\bibinfo {author} {\bibfnamefont {V.}~\bibnamefont
  {Bezchastnov}}\ and\ \bibinfo {author} {\bibfnamefont {T.}~\bibnamefont
  {Domratcheva}},\ }\bibfield  {title} {\bibinfo {title} {{Quantum-mechanical
  insights into the anisotropic response of the cryptochrome radical pair to a
  weak magnetic field}},\ }\href {https://doi.org/10.1063/5.0133943} {\bibfield
   {journal} {\bibinfo  {journal} {The Journal of Chemical Physics}\ }\textbf
  {\bibinfo {volume} {158}},\ \bibinfo {pages} {034303} (\bibinfo {year}
  {2023})}\BibitemShut {NoStop}%
\bibitem [{\citenamefont {Wiltschko}\ \emph {et~al.}(2010)\citenamefont
  {Wiltschko}, \citenamefont {Stapput}, \citenamefont {Thalau},\ and\
  \citenamefont {Wiltschko}}]{wiltschko2010directional}%
  \BibitemOpen
  \bibfield  {author} {\bibinfo {author} {\bibfnamefont {R.}~\bibnamefont
  {Wiltschko}}, \bibinfo {author} {\bibfnamefont {K.}~\bibnamefont {Stapput}},
  \bibinfo {author} {\bibfnamefont {P.}~\bibnamefont {Thalau}},\ and\ \bibinfo
  {author} {\bibfnamefont {W.}~\bibnamefont {Wiltschko}},\ }\bibfield  {title}
  {\bibinfo {title} {{Directional orientation of birds by the magnetic field
  under different light conditions}},\ }\href
  {http://doi.org/10.1098/rsif.2009.0367.focus} {\bibfield  {journal} {\bibinfo
   {journal} {Journal of The Royal Society Interface}\ }\textbf {\bibinfo
  {volume} {7}},\ \bibinfo {pages} {S163} (\bibinfo {year} {2010})}\BibitemShut
  {NoStop}%
\bibitem [{\citenamefont {Wiltschko}\ \emph {et~al.}(2006)\citenamefont
  {Wiltschko}, \citenamefont {Stapput}, \citenamefont {Thalau},\ and\
  \citenamefont {Wiltschko}}]{wiltschko2006avian}%
  \BibitemOpen
  \bibfield  {author} {\bibinfo {author} {\bibfnamefont {W.}~\bibnamefont
  {Wiltschko}}, \bibinfo {author} {\bibfnamefont {K.}~\bibnamefont {Stapput}},
  \bibinfo {author} {\bibfnamefont {P.}~\bibnamefont {Thalau}},\ and\ \bibinfo
  {author} {\bibfnamefont {R.}~\bibnamefont {Wiltschko}},\ }\bibfield  {title}
  {\bibinfo {title} {{Avian magnetic compass: fast adjustment to intensities
  outside the normal functional window}},\ }\href
  {https://link.springer.com/article/10.1007/s00114-006-0102-5} {\bibfield
  {journal} {\bibinfo  {journal} {Naturwissenschaften}\ }\textbf {\bibinfo
  {volume} {93}},\ \bibinfo {pages} {300} (\bibinfo {year} {2006})}\BibitemShut
  {NoStop}%
\bibitem [{\citenamefont {Nielsen}\ and\ \citenamefont
  {Chuang}(2010)}]{nielsen2010quantum}%
  \BibitemOpen
  \bibfield  {author} {\bibinfo {author} {\bibfnamefont {M.~A.}\ \bibnamefont
  {Nielsen}}\ and\ \bibinfo {author} {\bibfnamefont {I.~L.}\ \bibnamefont
  {Chuang}},\ }\href@noop {} {\emph {\bibinfo {title} {{Quantum computation and
  quantum information}}}}\ (\bibinfo  {publisher} {Cambridge university
  press},\ \bibinfo {year} {2010})\BibitemShut {NoStop}%
\bibitem [{\citenamefont {Cai}\ \emph {et~al.}(2012)\citenamefont {Cai},
  \citenamefont {Caruso},\ and\ \citenamefont {Plenio}}]{PhysRevA.85.040304}%
  \BibitemOpen
  \bibfield  {author} {\bibinfo {author} {\bibfnamefont {J.}~\bibnamefont
  {Cai}}, \bibinfo {author} {\bibfnamefont {F.}~\bibnamefont {Caruso}},\ and\
  \bibinfo {author} {\bibfnamefont {M.~B.}\ \bibnamefont {Plenio}},\ }\bibfield
   {title} {\bibinfo {title} {{Quantum limits for the magnetic sensitivity of a
  chemical compass}},\ }\href {https://doi.org/10.1103/PhysRevA.85.040304}
  {\bibfield  {journal} {\bibinfo  {journal} {Physical Review A}\ }\textbf
  {\bibinfo {volume} {85}},\ \bibinfo {pages} {040304} (\bibinfo {year}
  {2012})}\BibitemShut {NoStop}%
\bibitem [{\citenamefont {Schulten}\ \emph {et~al.}(1978)\citenamefont
  {Schulten}, \citenamefont {Swenberg},\ and\ \citenamefont
  {Weller}}]{schulten1978biomagnetic}%
  \BibitemOpen
  \bibfield  {author} {\bibinfo {author} {\bibfnamefont {K.}~\bibnamefont
  {Schulten}}, \bibinfo {author} {\bibfnamefont {C.~E.}\ \bibnamefont
  {Swenberg}},\ and\ \bibinfo {author} {\bibfnamefont {A.}~\bibnamefont
  {Weller}},\ }\bibfield  {title} {\bibinfo {title} {{A biomagnetic sensory
  mechanism based on magnetic field modulated coherent electron spin motion}},\
  }\href
  {https://www.degruyter.com/document/doi/10.1524/zpch.1978.111.1.001/html}
  {\bibfield  {journal} {\bibinfo  {journal} {Zeitschrift f{\"u}r Physikalische
  Chemie}\ }\textbf {\bibinfo {volume} {111}},\ \bibinfo {pages} {1} (\bibinfo
  {year} {1978})}\BibitemShut {NoStop}%
\bibitem [{\citenamefont {Johnsen}\ and\ \citenamefont
  {Lohmann}(2005)}]{johnsen2005physics}%
  \BibitemOpen
  \bibfield  {author} {\bibinfo {author} {\bibfnamefont {S.}~\bibnamefont
  {Johnsen}}\ and\ \bibinfo {author} {\bibfnamefont {K.~J.}\ \bibnamefont
  {Lohmann}},\ }\bibfield  {title} {\bibinfo {title} {{The physics and
  neurobiology of magnetoreception}},\ }\href
  {https://www.nature.com/articles/nrn1745} {\bibfield  {journal} {\bibinfo
  {journal} {Nature Reviews Neuroscience}\ }\textbf {\bibinfo {volume} {6}},\
  \bibinfo {pages} {703} (\bibinfo {year} {2005})}\BibitemShut {NoStop}%
\bibitem [{\citenamefont {Kominis}(2009)}]{kominis2009quantum}%
  \BibitemOpen
  \bibfield  {author} {\bibinfo {author} {\bibfnamefont {I.~K.}\ \bibnamefont
  {Kominis}},\ }\bibfield  {title} {\bibinfo {title} {{Quantum Zeno effect
  explains magnetic-sensitive radical-ion-pair reactions}},\ }\href
  {https://journals.aps.org/pre/abstract/10.1103/PhysRevE.80.056115} {\bibfield
   {journal} {\bibinfo  {journal} {Physical Review E}\ }\textbf {\bibinfo
  {volume} {80}},\ \bibinfo {pages} {056115} (\bibinfo {year}
  {2009})}\BibitemShut {NoStop}%
\bibitem [{\citenamefont {Jones}\ and\ \citenamefont
  {Hore}(2010)}]{jones2010spin}%
  \BibitemOpen
  \bibfield  {author} {\bibinfo {author} {\bibfnamefont {J.~A.}\ \bibnamefont
  {Jones}}\ and\ \bibinfo {author} {\bibfnamefont {P.~J.}\ \bibnamefont
  {Hore}},\ }\bibfield  {title} {\bibinfo {title} {{Spin-selective reactions of
  radical pairs act as quantum measurements}},\ }\href
  {https://www.sciencedirect.com/science/article/pii/S000926141000120X}
  {\bibfield  {journal} {\bibinfo  {journal} {Chemical Physics Letters}\
  }\textbf {\bibinfo {volume} {488}},\ \bibinfo {pages} {90} (\bibinfo {year}
  {2010})}\BibitemShut {NoStop}%
\end{thebibliography}%

\onecolumngrid
 

\section*{Supplementary material}

# Supplementary Information

Xiaoyu Chen,[1,2] Haibin Liu,[3, *] and Jianming Cai[1,2, †]

[1]*School of Physics, Hubei Key Laboratory of Gravitation and Quantum Physics,*

*Huazhong University of Science and Technology, Wuhan 430074, China*

[2]*International Joint Laboratory on Quantum Sensing and Quantum Metrology,*

*Institute for Quantum Science and Engineering,*

*Huazhong University of Science and Technology, Wuhan 430074, China*

[3]*School of Physics, Hubei University, Wuhan 430074, China*

(Dated: January 29, 2024)

**Rotation Matrix for Optimization and supplementary optimized hyperfine tensors**

In the main text, we plan to optimize the hyperfine tensor to make the spike sharper, so we show that $M = R_z(\gamma)R_y(\beta)R_x(\alpha)$, where $R_x(\alpha)$, $R_y(\beta)$ and $R_z(\gamma)$ represent rotations around the $\hat{x}(\hat{y}, \hat{z})$ by angle $\alpha(\beta, \gamma)$, so the matrix form of $M$ is as follows:

$$M = R_z(\gamma)R_y(\beta)R_x(\alpha) \tag{S.1}$$

$$= \begin{pmatrix} \cos(\beta)\cos(\gamma) & \cos(\gamma)\sin(\alpha)\sin(\beta) - \cos(\alpha)\sin(\gamma) & \cos(\alpha)\cos(\gamma)\sin(\beta) + \sin(\alpha)\sin(\gamma) \\ \cos(\beta)\sin(\gamma) & \cos(\alpha)\cos(\gamma) + \sin(\alpha)\sin(\beta)\sin(\gamma) & \cos(\alpha)\sin(\beta)\sin(\gamma) - \cos(\gamma)\sin(\alpha) \\ -\sin(\beta) & \cos(\beta)sin(\alpha) & \cos(\alpha)\cos(\beta) \end{pmatrix}.$$

The parameter $C$ we use in both the main text and the following section of this supplementary information for the optimized model systems is based on the parameters in Table SI.

**Efficient approach for radical pair's dynamics**

Radical pairs are involved in many chemical processes, in which each radical of a pair has an unpaired electron coupled to their surrounding nuclei and an external magnetic field via Hamiltonian. Usually, the two unpaired radical electrons are formed in an entangled state as excited by light, i.e., they begin in a singlet $|s\rangle$ or triplet state $|t_i\rangle$ ($i = 0, +, -$), and the reaction channels are spin state dependent. Therefore, an external magnetic field can affect the chemical reaction product, which is hypothesized to be connected with certain biological signal, and thus forms the

|  | Model system I | | | | | Model system II | | | | |
| --- | --- | --- | --- | --- | --- | --- | --- | --- | --- | --- |
| $k^{-1}(\mu s)$ | 10 | 8 | 5 | 2 | 1 | 10 | 8 | 5 | 2 | 1 |
| $A_{xx}$(mT) | -0.2678 | -0.1495 | 0.3622 | 0.3710 | 0.4070 | 0.0500 | -0.0499 | -0.0501 | -0.0501 | -0.0503 |
| $A_{yy}$(mT) | 0.0001 | 0.0002 | 0.0002 | -0.0002 | -0.0001 | 0.0006 | -0.0007 | 0.0000 | 0.0000 | 0.0000 |
| $A_{zz}$(mT) | -1.8172 | -0.8730 | 1.7458 | -1.1338 | 0.9330 | 0.0005 | -0.0003 | 0.0000 | 0.0000 | 0.0000 |
| $B_{xx}$(mT) | 1.8366 | 0.0150 | -1.7830 | -0.0002 | 1.0182 | -0.0002 | 0.0026 | -0.0300 | 0.0867 | 0.0000 |
| $B_{yy}$(mT) | -0.0305 | 0.0177 | 0.0103 | 0.0001 | 0.0001 | -0.0866 | 0.0004 | 0.0000 | 0.0000 | 0.0000 |
| $B_{zz}$(mT) | -0.0039 | -0.8854 | 0.0017 | 1.1932 | -0.0001 | -0.0002 | -0.0865 | -0.0058 | 0.0000 | -0.0869 |
| $\alpha$ | -1.0042 | 0.0189 | -0.1669 | 0.0266 | 0.1936 | 1.2461 | 0.2911 | -1.1853 | -0.0006 | 0.3101 |
| $\beta$ | -1.5521 | 3.1146 | -1.5867 | -0.0058 | 1.4998 | -2.2840 | -2.0492 | 1.2585 | -1.5776 | 0.8298 |
| $\gamma$ | -2.0299 | 1.8798 | 2.7745 | -1.3635 | -0.0089 | -1.1465 | 1.2512 | 0.3668 | 0.0002 | -1.1178 |

TABLE SI. The optimized hyperfine tensors for two different types of model systems with the radical pair lifetime $k^{-1} = 10$ $\mu$s, $8$ $\mu$s, $5$ $\mu$s, $2$ $\mu$s, $1$ $\mu$s respectively.

basis of a magnetic chemical compass [S1–S4]. The spin-selective recombination into the singlet and triplet product states can be described by the super operator

$$\mathcal{D}_u(\rho) = 2L_u\rho L_u^\dagger - L_u^\dagger L_u \rho - \rho L_u^\dagger L_u, \quad (S.2)$$

with $u = s, t_0, t_\pm$ for the singlet and triplet recombination channels respectively. The corresponding Lindblad operators are $L_s = |s,G\rangle\langle s,E|$, $L_0 = |t_0,G\rangle\langle t_0,E|$, $L_- = |t_-,G\rangle\langle t_-,E|$, $L_+ = |t_+,G\rangle\langle t_+,E|$, $|E\rangle$ and $|G\rangle$ indicate that radical pair are in the charge separated and recombined state respectively. Thus the radical pair's dynamics can be represented by

$$\dot{\rho} = -\frac{i}{\hbar}[H,\rho] - \sum_u k_u \mathcal{D}_u(\rho)/2, \quad (S.3)$$

where $k_u$ (usually we define $k_{t_0,t_\pm} \equiv k_t$) are the chemical reaction rate for different spin state. This master equation is strictly equivalent to the conventional phenomenological master equation in the charge separated subspace [S5, S6]:

$$\dot{\rho}_E = -\frac{i}{\hbar}[H,\rho_E] - \frac{k_s}{2}\{\rho_E, P_E^s\} - \frac{k_t}{2}\{\rho_E, P_E^t\}, \quad (S.4)$$

where $|E\rangle$ indicates that this equation happens in the charge separated subspace, $P_E^s$ and $P_E^t$ are the projectors into the singlet and triplet state individually.

Our aim is to calculate the time-dependent probability $p_s(t)$ that the radical pair is in a singlet state and hence the fraction of pairs ($\Phi_S$) that have recombined via the $|s\rangle$ state in the limit $t \to \infty$. This can be accomplished by solving master equation (S.3) or (S.4), which usually takes lots of computational resources. However, when $k_s = k_t \equiv k$, we can use a more efficient approach to do this. Using $P_E^s + P_E^t = I$, master equation (S.4) is simplified as (note that subscript $|E\rangle$ is dropped for convenience)

$$\dot{\rho} = -\frac{i}{\hbar}[H, \rho] - k\rho, \tag{S.5}$$

and the radical pair is assumed to be created in the singlet state, and the relevant nuclear spin is in the thermal equilibrium state, which is described by a identity matrix, so the initial state is

$$\rho_0 = \frac{1}{M}P^s, \tag{S.6}$$

where $M$ is the total number of nuclear spin configurations, $P^s = |s\rangle\langle s|$ is the singlet projection operator, then if we set $\rho = \rho_1 e^{-kt}$, we can derive that

$$\dot{\rho}_1 = -\frac{i}{\hbar}[H, \rho_1], \tag{S.7}$$

which describes a unitary process, so the solution of equation (S.5) takes the form

$$\rho(t) = e^{-\frac{i}{\hbar}Ht}\rho_0 e^{\frac{i}{\hbar}Ht}e^{-kt} = \frac{1}{M}e^{-\frac{i}{\hbar}Ht}P^s e^{\frac{i}{\hbar}Ht}e^{-kt}, \tag{S.8}$$

The time-dependent probability $p_s(t)$ that the radical pair is in a singlet state is [S7]

$$p_s(t) = \text{Tr}[\rho(t)P^s] = \frac{1}{M}\sum_{m=1}^{4M}\sum_{n=1}^{4M}|\langle m|P^s|n\rangle|^2 \cos[(\omega_m - \omega_n)t]e^{-kt}, \tag{S.9}$$

in which $|m\rangle$ and $|n\rangle$ are eigenstates of $H$ with energies $\hbar\omega_m$ and $\hbar\omega_n$, respectively. Hence the singlet product yield is

$$\Phi_s = k\int_0^\infty p_s(t)dt = \frac{1}{M}\sum_{m=1}^{4M}\sum_{n=1}^{4M}|P_{mn}^s|^2 \frac{k^2}{k^2 + (\omega_m - \omega_n)^2}, \tag{S.10}$$

where the 4M×4M terms are grouped into coherent terms with $m \neq n$ and incoherent terms with $m = n$.

### Incoherent terms and coherent terms cancelling out condition

In the main text, we have shown the contributions of incoherent and coherent terms to quantum needles with the optimized parameters corresponding to $k^{-1} = 10\ \mu$s of the model system II, Here,

we replenish the corresponding performance of the remaining parameters of the model system II. In Fig. S1, we can observe that except for small structures where the superposition of the contributions by the coherent terms and the incoherent terms can almost cancel each other, the contribution of the incoherent terms to the quantum needle of the singlet yield can be ignored. Overall, the directional precision in this newly found model system almost solely depends on the coherent terms.

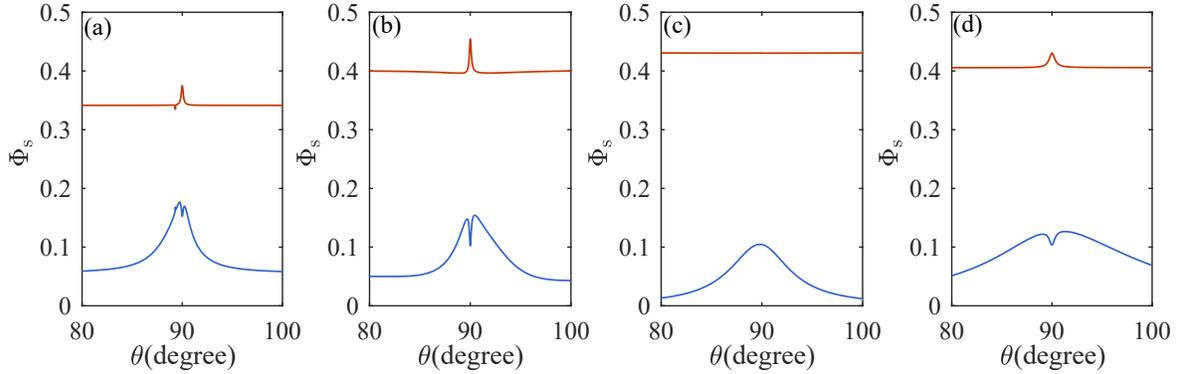

FIG. S1. The contributions of incoherent (red) and coherent terms (blue) to quantum needles with different radical pair lifetime $k^{-1} = 8$(a), $5$(b), $2$(c), $1$(d) $\mu$s and the corresponding optimized parameters $C$ are chosen from the model system II in Table SI.

When the contributions of the incoherent terms and coherent terms to the singlet yield are added up, we can see from Fig. 2 & Fig. 3 in the main text that some small needles can be completely canceled out. To understand this behavior, we analyze some typical cases here. For simplicity, we first choose Fig. 2(a) in the main text as an example. The contribution of coherent (or incoherent) terms to the singlet yield occurs while the eigenstates $|m\rangle$ and $|n\rangle$ of $H$ (see Eq. (1) in the main text) have avoided crossing. Thus, we demonstrate some typical avoid crossings and their contributions in Fig. S2. When the minimal gap between the avoided crossing energy-levels are much smaller than the reaction rate $k$, the contributions of coherent and incoherent terms are cancelled out, see Fig. S2(a)&(b). The same conclusion can also be reproduced with our optimized parameters. Some representative examples are presented in Fig. S3, and we will refrain from repeating the same points.

5

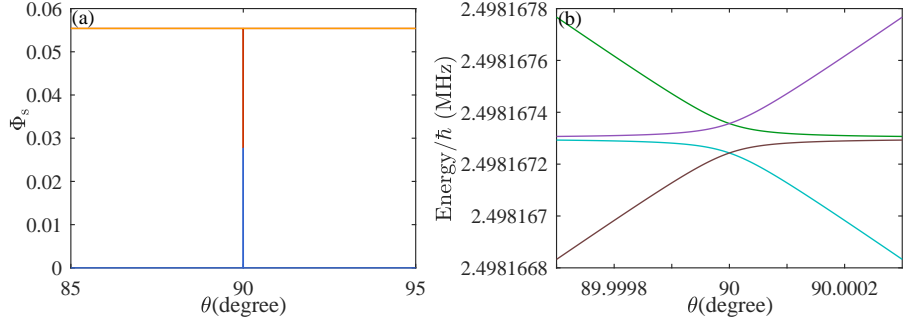

FIG. S2. (b) shows four energy levels with avoided crossings as functions of $\theta$. The contributions of corresponding incoherent terms (red), coherent terms (blue), and the sum of them (yellow) to the singlet yield are plotted in (a). The radical pair lifetime is set as $k^{-1} = 10$ $\mu$s and $C = C_0 = $ (-0.0989, -0.0989, 1.7569, 0.0, 0.0, 1.0812, 0, 0, 0) mT [S8].

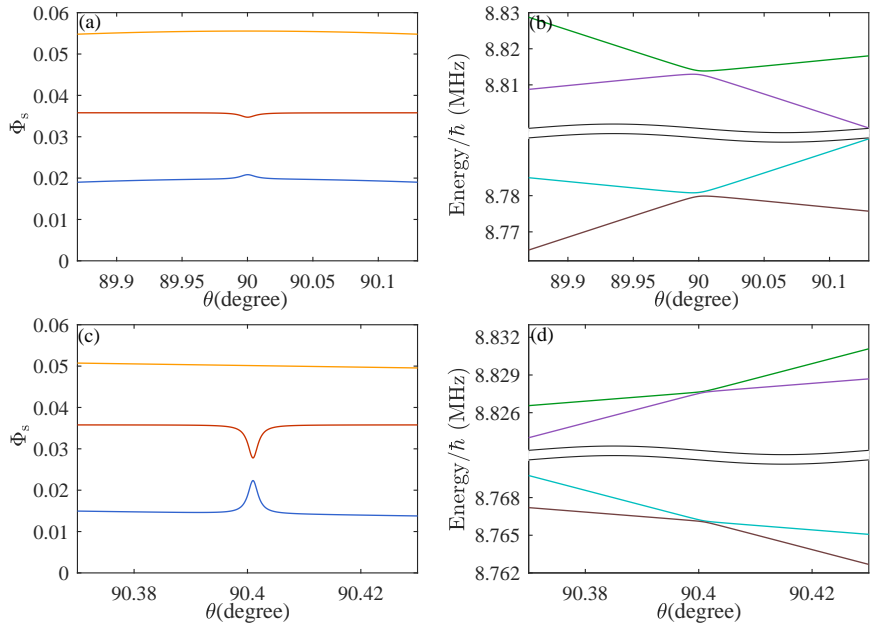

FIG. S3. (b) and (d) show four energy levels with avoided crossings as functions of $\theta$ around 90° and 90.4°, respectively. The contributions of corresponding incoherent terms (red), coherent terms (blue) and the sum of them (yellow) to the singlet yield are plotted in (a) & (c). The radical pair lifetime is set as $k^{-1} = 10$ $\mu$s and the corresponding optimized parameters $C$ are chosen from the model system II in Table SI.

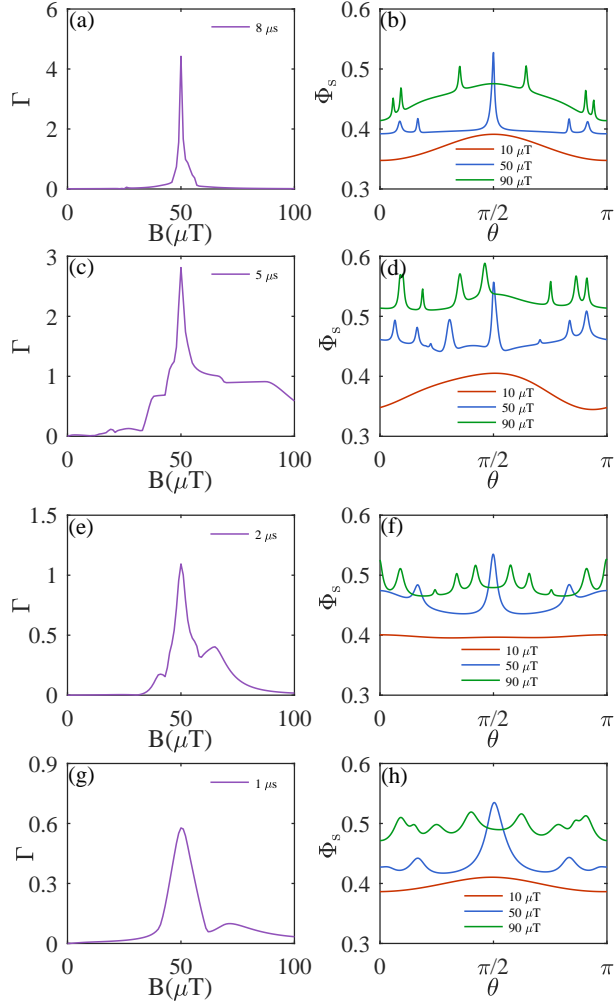

FIG. S4. The directional precision $\Gamma$ for the singlet yield as a function of the strength of the magnetic field $B$ in (a, c, e, g). (b, d, f, h) show the singlet yield $\Phi_s(\theta)$ for the magnetic field with $B = 10, 50, 90$ ($\mu$T), respectively. The radical pair lifetime $k^{-1} = 8, 5, 2, 1$ $\mu$s with corresponding optimized parameters $C$ in Table SI.

## Supplementary instructions for functional window

In the main text, we have shown the directional precision $\Gamma$ for the singlet yield as a function of the strength of the magnetic field $B$ and we also have shown the singlet yield $\Phi_s(\theta)$ for the magnetic field with $B = 10, 50, 90$ ($\mu$T) respectively, so here we replenish the corresponding performance of the remaining parameters of the model system II. In Fig. S4, although the value of $\Gamma$ is smaller than the main text, the functional window around the geomagnetic field persists.

* liuhb@hubu.edu.cn

† jianmingcai@hust.edu.cn

[S1] K. Schulten, C. E. Swenberg, and A. Weller, A biomagnetic sensory mechanism based on magnetic field modulated coherent electron spin motion, Zeitschrift für Physikalische Chemie **111**, 1 (1978).

[S2] T. Ritz, S. Adem, and K. Schulten, A model for photoreceptor-based magnetoreception in birds, Biophysical Journal **78**, 707 (2000).

[S3] S. Johnsen and K. J. Lohmann, The physics and neurobiology of magnetoreception, Nature Reviews Neuroscience **6**, 703 (2005).

[S4] C. T. Rodgers and P. J. Hore, Chemical magnetoreception in birds: the radical pair mechanism, Proceedings of the National Academy of Sciences **106**, 353 (2009).

[S5] I. K. Kominis, Quantum Zeno effect explains magnetic-sensitive radical-ion-pair reactions, Physical Review E **80**, 056115 (2009).

[S6] J. A. Jones and P. J. Hore, Spin-selective reactions of radical pairs act as quantum measurements, Chemical Physics Letters **488**, 90 (2010).

[S7] C. Timmel, U. Till, B. Brocklehurst, K. Mclauchlan, and P. Hore, Effects of weak magnetic fields on free radical recombination reactions, Molecular Physics **95**, 71 (1998).

[S8] H. G. Hiscock, S. Worster, D. R. Kattnig, C. Steers, Y. Jin, D. E. Manolopoulos, H. Mouritsen, and P. J. Hore, The quantum needle of the avian magnetic compass, Proceedings of the National Academy of Sciences **113**, 4634 (2016).


 
\end{document}